\documentclass[aps,preprint,12pt,floatfix,prc]{revtex4-1}
\usepackage{graphicx} 
\usepackage{amssymb}
\usepackage{amsmath}
\usepackage{bbm}
\usepackage{dsfont}
\usepackage{cases}
\usepackage{ulem}
\usepackage{braket}
\newcommand{\beq}{\begin{equation}}
\newcommand{\eeq}{\end{equation}}
\newcommand{\beqn}{\begin{eqnarray}}
\newcommand{\eeqn}{\end{eqnarray}}
\def \frac#1#2{ { #1 \over #2} }
\usepackage[capitalise]{cleveref}

\begin{document}
\title{A Dynamical Model of Light Halo Nuclei}
\author{Francisco Barranco$^{a}$, Gregory  Potel$^{a}$,
 Enrico Vigezzi$^{b}$}
\affiliation{$^{a)}$ Departamento de F\`isica Aplicada III,
Escuela T\'ecnica Superior de Ingenier\'ia, Universidad de Sevilla, Camino de los Descubrimientos, 	Sevilla, Spain}
\affiliation{$^{b)}$ INFN Sezione di Milano, Via Celoria 16, I-20133 Milano, Italy }



\begin{abstract}
We present a review of theoretical studies of the structure and reactions of  N=7 and N=8 nuclei in the vicinity of 
$^{11}$Li, carried out within a  framework based on Nuclear Field Theory. The coupling of valence nucleons to low-lying surface vibrations of the spherical core plays a central role, giving rise to self-energy processes that renormalize single-particle states and transfer form factors, as well as to an induced pairing interaction arising from  the exchange of collective vibrations, which renormalizes the bare pairing force.
Excitation spectra and cross sections for one- and two-nucleon transfer reactions populating states in the quasi-continuum are calculated and compared with available experimental data. Collective excitations in the particle-particle channel are  investigated, with particular emphasis on Giant Pairing Vibrations and on their damping mechanisms arising from coupling to more complex configurations and continuum states.  Comparisons with other theoretical  schemes  are also presented. We conclude that a coherent understanding of  experimental data requires the detailed consideration of  particle-vibration coupling effects.
\end{abstract}

\maketitle

\section{Introduction}

The discovery of the neutron halo in 
$^{11}$Li forty years ago, celebrated at this Conference, represented a major breakthrough in the exploration of the limits of nuclear stability. Since then, many experimental properties of 
$^{11}$Li and of neighboring nuclei have been firmly established. The interpretation of these data and the development of theoretical models capable of providing a unified quantitative description of both structure and reaction observables in this region of the nuclear chart, remain the subject of active debate \cite{sagawa_review1,Moro20251}.

Halo nuclei are characterized by strong renormalization effects with respect to a simple mean-field picture. In particular, the weak binding of the valence nucleons and the proximity of the continuum enhance the role of correlations and of the coupling between single-particle and collective degrees of freedom
\cite{Broglia_exo1}. 
Two-body models (for one-neutron halos) and three-body models (for two-neutron halos), based on a frozen core (spherical or deformed), are often able to reproduce several observables by introducing specific parameters, for example parity-dependent potentials. However, such approaches may encounter difficulties when dealing with phenomena that involve the dynamics of the core. The present contribution aims at clarifying the role of the coupling between core and valence degrees of freedom, emphasizing in particular the importance of surface fluctuations, whose relevance has been highlighted by several key experiments. We review a number of studies we have carried out on the structure and reactions of  N=7 and N=8 nuclei around $^{11}$Li.
Throughout the paper, we emphasize the importance of checking the structure theory against one- and two-neutron transfer cross sections, as they are one of the main tools to probe single-particle spectroscopic factors and pairing correlations, together with breakup and (p,pn) reactions.

Our theoretical framework is based on Nuclear Field Theory (NFT) \cite{NFT}, which provides a consistent treatment of the nuclear many-body problem through the coupling of single-particle and collective degrees of freedom. In particular, NFT incorporates ground-state correlations, which are essential for understanding the renormalization of single-particle energies, as well as the exchange of collective modes that renormalize the pairing interaction in two-neutron halos and contribute significantly to the stability of $^{11}$Li.
In traditional approaches to particle-vibration coupling (PVC) in heavier systems, it is usually sufficient to consider the coupling with the vibrations of the well-bound core. In the case of  $^{11}$Li, however, the proximity of the continuum leads to the appearance of substantial low-energy dipole strength and thus to a very large dipole polarizability, which is not present in the $^9$Li core. 
The coupling to this low-lying dipole strength must therefore be taken into account (bootstrap effects), paying particular attention to strong Pauli constraints, since part of the dipole strength is associated with low-lying orbitals and should not be counted twice.

Our approach is particularly suitable for the study of collective pairing vibrations, including effects beyond the traditional particle-particle RPA. In the final part of this contribution we show that such effects can lead to significant modifications of the spectrum calculated within pp-RPA, especially for the high-lying collective modes, the so-called Giant Pairing Vibrations. These modes were predicted within the pp-RPA framework but have not yet been clearly observed experimentally. We discuss in this context a recent two-neutron transfer experiment populating  $^{14}$C, that observed  a bump in the excitation spectrum, that was  interpreted as a possible signature of a Giant Pairing Vibration.

In the course of this paper we also briefly discuss analogies and differences between our approach and other theoretical models that treat the coupling between valence and core particles in different ways. The present discussion represents only a glimpse of the variety of approaches that have been developed to study halo systems; in particular, a comparison with cluster models is beyond the scope of this work.

The paper is organized as follows. In chapters 2 and 3 we outline the formalism for the A+1 and A+2 systems respectively, taking into account the effects of the coupling between particles and surface vibrations. In chapter 4 applications to $^{11}$Be, $^{10}$Li, $^{11}$Li and $^{14}C$ are presented, and  the formalism of second order DWBA adopted for the calculation of  (p,t)  cross sections is briefly reviewed. Finally, in chapter 5 conclusions are formulated.

\section{Revisiting the Particle-Vibration Coupling in A+1 systems}

The Hamiltonian of the system is defined as
\begin{equation}
H_{1\nu} = H_0 + H_{vib} + H_{PVC}
\end{equation}
where $H_0 = K + V $ denotes its single-particle part
and $H_{PVC}$ is the term that couples single-particle and vibrational degrees of freedom as discussed below. We assume spherical symmetry. The Hamiltonian $H_{vib}$ describes the harmonic vibrations of the core and is given by
\begin{equation}
H_{vib} = \sum_{LM} \hbar\omega_{L} [\Gamma_{LM}^{\dagger}\Gamma_{LM} + 1/2],
\end{equation}
where the operators $\Gamma^{\dagger}_{LM}$ and $\Gamma_{LM}$ create a phonon of  angular momentum $L$, component on the $z$-axis $M$ and energy  $\hbar\omega_{L}$,  $\Gamma^\dagger_{LM}\ket{0} = \ket{\Phi_{LM}}$.

 The determination of the Hamiltonian $H_{1\nu}$ could be based on a single effective interaction, which can be used to produce a mean field with a Hartree-Fock (HF)  calculation,  a spectrum of vibrational excitations based on the Random Phase Approximation (RPA) and the matrix elements of $H_{PVC}$ \cite{Colo_Bort}. This approach is very attractive, but it will not be followed here because of 
the following considerations:

a)  The parameters of the most widely used effective interactions (namely of Skyrme and Gogny type in the non relativistic framework) have been at least partially fitted on experimental  properties, that will be modified by PVC effects. This can lead to double counting, because refitting the interaction including PVC effects is in practice computationally too demanding.   

b) Single-particle properties and continuum thresholds in light nuclei are often not accurately reproduced by  effective interactions that are built  to provide a global description of nuclei   over the  whole nuclear table. This fact prevents a quantitative description of low-energy excitations in the continuum, in particular for weakly bound  nuclei. 

c) Even if this approach has the advantage of self-consistency, it is still based on the choice of a particular interaction, typically based on a dozen of parameters. 

We will instead prefer to determine the few parameters defining a Woods-Saxon potential $V(r)$ by fitting single-particle properties {\bf including} the effects of PVC, calculated making use of the experimental properties (energies $\hbar \omega_{L}$ and deformation parameters $\beta_{L}$) of the low-lying collective vibrations.

The use of a single-particle basis based on a Woods-Saxon potential has some advantages as compared with other standard approaches based on harmonic oscillator wavefunctions. The use of a finite potential well ensures a fair reproduction of the appropriate asymptotic behavior, both for bound (exponential decrease) and unbound (oscillatory behavior) states. This has a considerable impact on the ability to reproduce experimental one- and two-particle transfer cross sections populating bound states (see Sect. \ref{S4}), and also allows for a straightforward calculation of the phase shifts associated with unbound neutron states in the experimentally observed channels. The continuum is discretized by performing the calculations in a finite spherical box with a large enough radius, checked for convergence. By changing the box radius, it is possible to scan the phase shifts above the particle threshold as a function of the energy in an essentially continuous fashion. The energy derivative of the phase shifts then provides the level density (spectral function), from which the widths and energies of resonances can be extracted  (See the insets in Fig \ref{fig:levels_11Be}).

The dynamical interaction between single-particle motion and the core is then described by the usual  particle-vibration term $H_{PVC}$ \cite{BohrII}:
\begin{equation}
H_{PVC}= \sum_{LM} -r \frac{dV(r)}{dr} \frac{\beta_{L}}{\sqrt{2L+1}} Y_{LM}(\Omega) [\Gamma_{LM}^{\dagger} + (-1)^{M}\Gamma_{L-M}]
\end{equation}

We will limit ourselves to configurations including at most one phonon, with the exception of the case of $^{11}$Be discussed below. The Hamiltonian matrix of the model  can then be written schematically as in  Table \ref{Table_H1nu}.

\begin{table}[h!]
\begin{center}
\begin{tabular}{|c|c  | c | c |c|}
\hline
$H_{1v}$ & p & p$\otimes b$&  h$\otimes b$  \\ 
\hline
 p&  $H_{mf} $ & $H_{PVC}$ & $H_{PVC}$ \\ 
 \hline
p$\otimes$b & $H_{PVC}$ & $H_{mf}$  +  $H_{vib}$ &   0  \\ 
\hline
h$\otimes$b& $H_{PVC}$ & 0                                     & $H_{mf}$ - $H_{vib}$  \\ 
\hline
\end{tabular}
\caption{\footnotesize  Schematic representation of the $H_{1\nu}$ matrix, where $p$ and $h$ represent a particle and a hole state associated with the   mean field $H_{mf}$, while $b$ represents a phonon state.}
\label{Table_H1nu}
\end{center}
\end{table}	

The minus sign in the diagonal term $h \otimes b$ is related to the
 so called "anomalous coupling" to hole states accounting for the ground state correlations present in the A (core) subsystem.

The eigenvalues of the A+1 system are those lying above the hole states.
The other eigenvalues are interpreted as associated to states of the A-1 system.

This eigenvalue problem may be reformulated as an energy-dependent problem  by projection onto the pure fermion (p) sector by means of the Bloch-Horowitz procedure \footnote{The Bloch-Horowitz (BH) projection provides a rigorous framework for mapping the full Hilbert space onto a manageable model subspace (\(P\)). By constructing an energy-dependent effective Hamiltonian, the method accounts for the influence of the excluded space (\(Q\)) while reducing the dimensionality of the many-body problem.

}.

This projection allows a re-formulation of the effects of the coupling to vibrational states by using the well known and rather intuitive language of the Feynman diagrams used in Quantum electrodynamics, and in general in Quantum Field Theories, which 
also applies to  NFT. 
Note that the Feynman-diagram formalism is required for a proper treatment within NFT of higher-order effects involving multiple phonons, since some of the NFT rules are formulated in terms of the topology of the diagrams.

\begin{figure}
    \centering
    \includegraphics[width=0.5\linewidth]{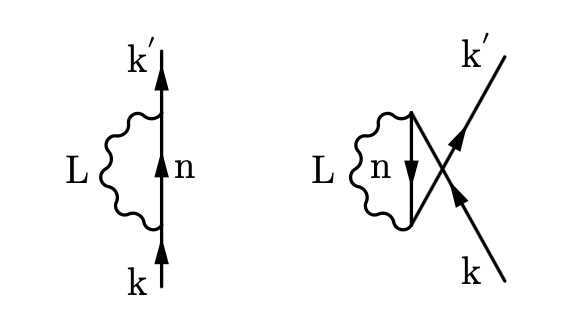}
    \caption{\footnotesize Polarization (left) and correlation (right) self-energy diagrams.}
    \label{fig:selfen}
\end{figure}

In the present case , the BH-projection leads to the well-known self-energy diagrams, which are those at the basis of the Lamb-shift in QED (see Fig.\ref{fig:selfen}). In fact, performing the BH projection onto the pure particle space, the self-energy $\Sigma$-term appears as an additional energy-dependent correction to the `pure single-particle Hamiltonian, as shown in Table \ref{Table_H2nu}.

\begin{table}[h!]
\begin{center}
\begin{tabular}{|c|c  | c | c |c|}
\hline
$H_{1v} (\omega) $ & p  \\ \hline
 p &  $H_{mf}$ + $\Sigma(\omega)  $\\ \hline
\end{tabular}
\caption{\footnotesize Schematic representation of the energy-dependent $H_{1\nu}$ matrix.}
\label{Table_H2nu}
\end{center}
\end{table}

 The total self-energy matrix element between a nucleon in state $k$ and state $k'$ is expressed as the sum of two contributions diagrammatically shown in Fig. \ref{fig:selfen}:
  
\begin{equation}
\Sigma(k, k',  \omega) = \Sigma_{pol}(k, k',\omega) + \Sigma_{corr}(k,k',\omega)
\end{equation}
where $k$ stands for ($n, l, j, m$) and $k'$ for ($n', l, j, m$). Note that $l,j, m$ are conserved.

The polarization process (obtained  projecting the  $p \otimes b$ configurations) describes the emission and reabsorption of a phonon with energy $\hbar \omega_L$. The matrix element of the self-energy $\Sigma_{pol}(k, k',  \omega)$ is determined according to  the rules for the calculation of Feynman diagrams.
Namely, the product of the two vertices $<k|H_{PVC}|n,L>$ and $<n,L|H_{PVC}|k'>$ is divided by the 
energy difference between the energy of the state $\hbar \omega$ and the energy of the intermediate state
$\epsilon_n + \hbar \omega_L$:
\begin{equation}
\Sigma_{pol}(k, k',  \omega) = \sum_{p_n, b_{LM}} \frac{\langle k | H_{PVC} | p_n, b_{LM} \rangle\langle p_n, b_{LM} | H_{PVC} | k' \rangle}{\hbar \omega - (\epsilon_n + \hbar \omega_L) + i\eta}
\end{equation}

By projecting the anomalous $h \otimes b $ components, one obtains the correlation term:
\begin{equation}
\Sigma_{corr}(k, k',  \omega) = \sum_{h_n, b_{LM}} \frac{\langle k | H_{PVC} | h_n, b_{LM} \rangle\langle h_n, b_{LM} | H_{PVC} | k' \rangle}{\hbar \omega - (\epsilon_n - \hbar \omega_L) + i\eta},
\end{equation}
where now $h_n$ denotes a hole in the intermediate state.
This term is an essential exchange correction, that accounts for the fact that the valence neutron inhibits the  shape fluctuations (virtual excitations) of  the core, resulting in a partial increase in the energy of the system. This is apparent in the form of the corresponding diagram in Fig. \ref{fig:selfen} , in which the intermediate virtual state, when $k$ and $k'$ are equal, represents the Pauli forbidden configurations. The value of the diagram  is equal to the  core correlation energy that must be eliminated, in keeping with the minus sign in the above expression, related to  the crossing of the $k$
and the $k'$ fermion lines in the diagram.

The matrix element of $H_{PVC}$ is given by:
\begin{equation}
\langle p_n, b_{LM} | H_{PVC} | k \rangle = - \frac{\beta_L}{\sqrt{2L+1}} \int \phi_n(\vec{r}) r \frac{dV(r)}{dr} Y_{LM}(\hat{r}) \phi_k(\vec{r}) d^3r
\label{hpvc_sep}
\end{equation}
One could use a more microscopic approach,   adopting an effective  two-body interaction $v(r,r') $ and
calculating the matrix elements in the Random Phase Approximation (RPA): 
\begin{equation}
\langle p_n, b_{LM} | H_{PVC} | k \rangle = \int \phi_n (\vec{r}) \rho^{RPA}_{LM}(\vec{r}') \phi_k(\vec{r}) v(\vec{r},\vec{r}’) d^3r d^3r’,
\label{hpvc_rpa}
\end{equation}
where $v(\vec r, \vec r’)$ is the two-body interaction used in the RPA calculation,
with the associated  transition density 
\begin{equation}
\rho^{RPA}_{LM}(\vec{r}) = \sum_{ph} [X_{ph}(LM) + Y_{ph}(LM)] [\phi_p(\vec{r}) \phi_h(\vec{r})]_{LM},
\label{trans_rpa}
\end{equation}
where $X_{ph}$ and $Y_{ph}$ denote the forward and backward RPA amplitudes respectively.
Assuming a separable interaction, Eq. (\ref{hpvc_rpa}) reduces to Eq. (\ref{hpvc_sep}) 
(see \cite{Idini} for more  details.)

The normalization of a quasiparticle wave function of energy $E$ is given by:
\begin{equation}
\sum_{k,k'} \left( \delta_{k,k'} - \frac{\partial \Sigma(k, k' \omega)}{\partial \omega} \right)_{\omega = E} \xi_k(E) \xi_{k'}(E) = 1
\end{equation}
where $\xi_k$ are the amplitudes of a given eigenstate on the single particle $k$. This expression accounts for the energy-dependence (existence of Bloch-Horowitz hidden components) of the self-energy. As a consequence the quasiparticle content is less than 1.

In the simple case in which  only one $n,l,j,m$ state is present in the "explicit" basis, the above equation reduces to

\begin{equation}
  \xi_k(E) =  \left( 1 - \frac{\partial \Sigma(k, \omega)}{\partial \omega} \right)_{\omega = E}^{-1}
\end{equation}
This is usually referred to as  the $Z_k$ strength of the quasi-particle.



So far we have accounted for the presence of at most one surface phonon in the model basis. For large values of the  deformation parameters, it may occur that the presence of two-phonon configurations become relevant. In such cases one needs to go to the next order in the PVC/NFT Feynman-diagrammatic expansion, and as a consequence new phenomena are taken into account, leading to the  renormalization of vertices and  phonon energies. Also, rainbow type diagrams must be added  to the self-energy.

\begin{figure}
    \centering
\includegraphics[width=1.\linewidth]{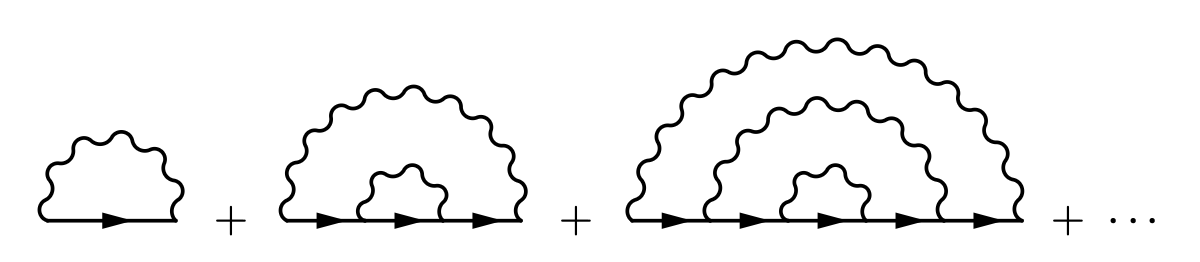}
    \caption{\footnotesize The rainbow series.}
    \label{fig:rainbows}
\end{figure}

 A rainbow diagram is similar to that of the polarization self-energy already seen above, but with  several phonons being emitted and reabsorbed, as shown in Fig. \ref{fig:rainbows} in the case of two and three phonons  (the diagrams are shown horizontally for pictorial convenience). Adding further phonons the so called rainbow series is constructed. This series is convergent and may be calculated (see below).

\begin{figure}
    \centering
\includegraphics[width=0.5\linewidth]{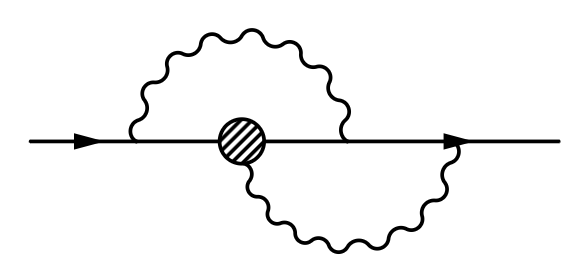}
    \caption{\footnotesize The vertex renormalization.}
    \label{fig:vertex}
\end{figure}

 The lowest  order  diagram associated with  vertex renormalization is similar to the two-arc rainbow but for the fact that the two phonon lines cross so that the first emitted phonon is reabsorbed before the second one. So, the first emitted phonon appears and disappears around the first vertex of the second phonon, and this vertex is said to be renormalized 
 (see Fig. \ref{fig:vertex}). Higher order vertex renormalization diagrams would imply more phonons around the vertex. The net effect of the vertex renormalization series would be a modification of the deformation parameters entering the self-energy diagrams, with respect to  the RPA value. 
 In general, all the  properties of the phonons besides  the deformation parameters are renormalized,  such as energy, quadrupole moment, etc.

Vertex and phonon energy renormalization do already occur in the A (core) system, so that when using experimental data for $\beta_L$ and  $\omega_L$ they are  effectively included. Instead, rainbow diagrams (see Fig. \ref{fig:rainbows}) are specific of the A+1 system and should be included explicitly.
A simple way to include them is to iterate the energy-independent eigenvalue problem, using at each iteration the (many) quasiparticle states of the previous iteration as "pure" fermions of reduced strength $Z_k$. The number of surface phonons included in the rainbow diagrams is equal to the number of iterations. This iterative procedure may be easily performed many times, although the number of "pure" states and  the matrix size rapidly increase. Finally, computing power would put a limit, but in the cases we have studied, adequate  convergence can be reached already after two or three iterations.  If necessary, some statistical coarse grain averaging can be used to approximately account for the infinite rainbow series.\cite{OurVenice, Soma_PRC89, Muther_NPA581}.

Another relevant process that appears by including the coupling to  two or more phonons is the phonon-phonon interaction.

\begin{figure}[t!]
    \centering
    \includegraphics[width=0.5\linewidth]{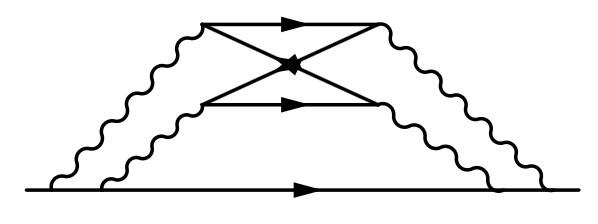}
    \caption{\footnotesize The butterfly diagram.}
    \label{fig:farfalla}
\end{figure}

If  phonons were  perfect bosons, they could coexist without disturbing each other. But nuclear vibrations are only approximately bosonic. At the microscopic level they are linear combinations of p-h configurations. Thus, when two phonons appear simultaneously in a diagram like the two-arc rainbow, from time to time  they occupy the same p-h configuration, violating  the Pauli principle. 
This violation should be corrected, because the Pauli principle reduces the available phase space, and correspondingly also  the correlation energy  calculated with the rainbow diagram. This reduction can be taken into account  in NFT by including the so-called butterfly diagram  named after its characteristic wing-like shape, see Fig. \ref{fig:farfalla} 
in which the two phonons  exchange the particle (or hole) that is common to both configurations and therefore Pauli forbidden.
The butterfly diagram amounts to  an energy contribution that reduces  the  correlation energy calculated with the two-arc rainbow, and is of the same order as the three-arc rainbow, which must also be included in a consistent calculation of the correlation energy.
See below for more discussion on this point in the case of $^{11}$Be.

\section{The Particle-Vibration Coupling in A+2 systems}

In this Section we outline our  theoretical framework to compute the $I^{\pi}=0^{+}$ pair addition (or removal) modes of a nucleus with A+2 (or A-2) nucleons, built on top of a spherical, vibrating $A$ core.  We will extend the previous PVC scheme to the case of two valence neutrons interacting through a pairing force $V_{int}$.   We assume 
that the core is not superfluid, although the theory can be extended to deal with open shell nuclei \cite{Idini}. The excitation spectrum  includes both low-lying bound Pairing Vibrations (PV) and the predicted high-lying Giant Pairing Vibrations (GPV) in the continuum (see section 4.4 below).
We notice that we aim at an interpretation of the excitation spectrum of the system, and not of its bulk properties, like the total binding energies, which are better studied with other approaches.

The Hamiltonian of the system is thus defined as
\begin{equation}
H_{2\nu} = H(1) + H(2) + V_{int}(1,2) + H_{vib}
\end{equation}
where $H(i) = K(i) + V(i) + H_{PVC}(i)$ for $i=1,2$. 


An essential consistency requirement is a good description of the A+1 and A-1 systems, that was described above.

We will explicitly describe the pair addition case. The basis states for the $A+2$ system include the following types of configurations:
\begin{itemize}
    \item \textbf{i) fermion pp:} $|[n_{lj}n'_{lj}]_{0^+}\rangle$ ($e_{nlj}, e_{n'lj} > E_F$) 
    \item \textbf{ii) fermion pp boson:} $|[[n_{lj}n'_{l'j'}]_{L} \otimes \Phi_{L}]_{0^+}\rangle$ ($e_{nlj}, e_{n'lj} > E_F$) 
    \item \textbf{iii) fermion ph boson:} $|[[n_{lj}n'_{l'j'}]_{L} \otimes \Phi_{L}]_{0^+}\rangle$ ($e_{nlj} < E_F, e_{n'lj} > E_F$)
    \item \textbf{iv) fermion hh:} $|[n_{lj}n'_{lj}]_{0^+}\rangle$ ($e_{nlj}, e_{n'lj} < E_F$)
\end{itemize}

We will mostly   utilize  a finite range monopole  force  restricted to the $pp$-sector for the  pairing interaction $V_{int}$. In Section 4 we will also consider the effects of a quadrupole interaction acting between $pp \otimes  b$ configurations.

The resulting Hamiltonian matrix (see  Table \ref{Table_Ext_RPA}) is an extended pp-RPA matrix. If the deformation parameters $\beta_{\lambda}$  are set to zero, the standard pp-RPA is recovered \cite{Blanchon,Ring}. 
Note that we will focus on the PV addition states, and we will not need a detailed description of the hole states. For this reason we ignore their coupling with $ph \otimes b$  configurations, and we do not include a $hh \otimes  b $ sector in $H_{2\nu}$. 

\begin{table}[h!]
\begin{center}
\begin{tabular}{|c|c  | c | c |c|}
\hline
$H_{2v}$ & pp & pp$\otimes b$&  ph$\otimes b$& hh   \\ \hline
 pp&  $H_{mf} + V_{int}$ & $H_{PVC}$ & $H_{PVC}$& $V_{int}$ \\ \hline
pp$\otimes$b & $H_{PVC}$ & $H_{mf}$  +  $H_{vib}$ + $V_{int}$ &   0 & 0  \\ \hline
ph$\otimes$b& $H_{PVC}$ & 0                                     & $H_{mf}$ - $H_{vib}$ + $V_{int}$  & 0 \\ \hline
hh & $-V_{int}$ & 0  & 0&  $H_{mf} + V_{int}$ \\
\hline
\end{tabular}
\caption{\footnotesize Schematic representation of the extended pp-RPA matrix. The sub-matrix connecting pp-states  among themselves is the conventional 
particle-particle Tamm-Dancoff Approximation, or pp-TDA matrix.
The sub-matrix connecting pp- and hh- states is the conventional pp-RPA matrix. }
\label{Table_Ext_RPA}
\end{center}
\end{table}

The wavefunctions of the A+2 systems can be used to compute the cross sections of two-nucleon transfer reactions. In some of the  following examples, we will  limit ourselves to the calculation of monopole strength functions. In the case of pair addition, the latter are defined as 
\begin{equation}
\begin{aligned}
S_k(R_{box}) = & |\sum_{pp'} X(k)_{pp'} \int dr \psi_p(r) \psi_{p'}(r) f(r) \langle j_p || Y_0 || j_{p'} \rangle + \\
& \sum_{hh'} Y(k)_{hh'} \int dr \psi_h(r) \psi_{h'}(r) f(r) \langle j_h || Y_0 || j_{h'} \rangle |^2
\end{aligned}
\label{eq:strength}
\end{equation}
 where $f(r) = dV/dr$ is the radial form factor and X's and Y's are the forward and backward amplitudes of our extended pp-RPA. Our calculations are carried out using a box of a given radius $R_{box}$. In order to avoid spurious continuum discretization effects, we make a series of calculations changing the box size, and then take the average 
over the different strength functions. In the no coupling limit ($\beta_{\lambda}$ =0),
we have verified \cite{prl_gpv} that the results of the exact continuum pp-RPA \cite{Matsuo} are accurately reproduced  by this procedure .  

\subsection{The pairing induced interaction}

An alternative, though mathematically equivalent, view of the effects produced by the PVC as exposed in the previous paragraphs, and often found in literature, is through the concepts of self-energy and of pairing induced interaction.

\begin{figure}
    \centering
    \includegraphics[width=0.7\linewidth]{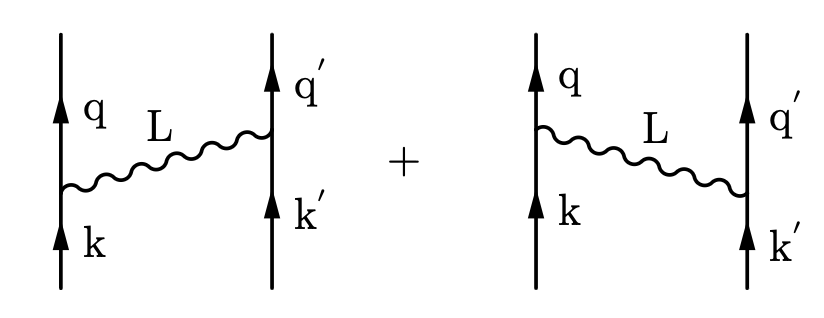}
    \caption{\footnotesize Feynman diagrams representing the induced interaction processes.}
  \label{fig:induced}
\end{figure}

The self-energy has been introduced in the preceding Section, and here appears twice since we are dealing with two neutrons.
If $H_{PVC}(1)$ is utilized in both emission and absorption, then only the first neutron comes into play (the second being a passive observer, implying $q' = k$'), and the self-energy already discussed in the previous chapter appears for the first neutron. A similar term appears for the second neutron when $H_{PVC}(2)$ is used twice.
A new type of term  appears when performing the Bloch-Horowitz reduction to pure fermionic states, namely  the interaction mediated by the exchange of surface vibration quanta. This term is the counterpart of the exchange of virtual photons between electrons appearing in QED.
In fact, assuming a monopole pairing interaction, so that there are no $V_{int}$ terms coupling the  $pp \otimes b$ configurations among themselves, one obtains  the Bloch-Horowitz effective Hamiltonian matrix element
\begin{multline}
M_{\mathrm{BH}}(k,k'; q,q'; \omega) =  \\
\sum_{n,m,L}
\frac{
\langle k k' | [H_{PVC}(1)+H_{PVC}(2)] | n m, L \rangle
\langle n m, L | [H_{PVC}(1)+H_{PVC}(2)] | q q' \rangle
}{
\omega - (\epsilon_n + \epsilon_m \pm \hbar \omega_L) 
}
\end{multline}
where the minus sign before $\hbar \omega_L$ applies when $n$ or $m$ is a hole state.
Then there appears the possibility of going through $H_{PVC}(1)$ in the first place, what implies $m=q'$, and through $H_{PVC}(2)$ in the second factor, what implies $n=k$, and vice versa, meaning that a surface quanta is emitted by the first neutron and subsequently absorbed by the second (see a diagrammatic representation of these two processes in Fig. \ref{fig:induced}). This kind of contribution is thus defined as the induced pairing matrix elements or $V_{ind}$. Note that in this kind of process both $k$ and $k'$ indices may change into $q,q'$, since both neutrons participate actively.

The detailed expression of  the matrix element of $V_{ind}$ in the important case of  pairs coupled to $J=0^+$ is given by 
\begin{eqnarray}
V_{ind}(k,k';q,q';\omega) = 
\sum_{L} \frac{<j_{k}\frac{1}{2}j_q \frac{1}{2}|L0>^2}{8\pi(2L+1)} 
\frac{\beta_L^2  <j_{k}|r dV/dr|j_q><j_{k'}|r dV/dr|j_{q'}>}{\omega - (\epsilon_{k'}+ \epsilon_q + \hbar \omega_L)} + \notag & \\
\sum_{L} \frac{<j_{k}\frac{1}{2}j_q \frac{1}{2}|L0>^2}{8\pi(2L+1)} 
\frac{\beta_L^2  <j_{k}|r dV/dr|j_q><j_{k'}|r dV/dr|j_{q'}>}{\omega - (\epsilon_{k}+ \epsilon_{q'} + \hbar \omega_L)}\end{eqnarray}   


Thus the effective Bloch-Horowitz Hamiltonian matrix elements can be written as the sum of self-energy terms and pairing induced interaction term
\begin{equation}M_{\mathrm{BH}}(k,k'; q,q'; \omega) =  \\
\Sigma_1(k, q; \omega) \delta(k',q')+ \Sigma_2(k', q'; \omega) \delta(k,q) + V_{ind}(k, k'; q, q';  \omega),
\end{equation}
where the term $V_{ind}(k,k',q,q';\omega)$ is analogous to the attraction between two electrons through the exchange of phonons from the lattice  appearing in the BCS theory of metals \cite{Schrieffer}. The essential difference is, of course, that in the nuclear case the phonons represent excitations produced by the nuclear systems itself.  The exchange of collective vibrations of the mean field (nuclear phonons), such as quadrupole ($2^+$) or octupole ($3^-$) modes contributes to nuclear pairing \cite{Idini}, in addition to  the bare attractive pairing interaction $V_{int}$, which has no counterpart in the case of metals.

Thus, after Bloch-Horowitz projection we obtain the energy-dependent RPA matrix shown in 
{Table \ref{Table_Ext_RPA_1}.
\begin{table}[h!]
\begin{center}
\begin{tabular}{|c|c  | c | c |c|}
\hline
$H_{2v}(\omega$) & pp & hh   \\ \hline
 pp &  $H_{mf}$  + $\Sigma_1(\omega)$ + $\Sigma_2(\omega)$  + $V_{int}$+$V_{ind}(\omega)$ & $V_{int}$ \\ \hline
hh & $-V_{int}$ & $H_{mf} + V_{int}$ \\
\hline
\end{tabular}
\caption{\footnotesize Schematic representation of the energy dependent pp-RPA matrix for the A+2 system after Bloch-Horowitz projection. }
\label{Table_Ext_RPA_1}
\end{center}
\end{table}	
}

\section{Applications to N=7 and N=8 isotones}\label{S4}

We have studied the spectra and reactions of Li (Z=3), Be (Z=4) and C (Z=6) nuclei having 7 and 8 neutrons. 
In particular the nucleus $^{11}$Be has become the prototype of a single-neutron halo, just as $^{11}$Li is probably the most studied case of two-neutron halo.  In turn, recent transfer experiments between heavy ions populating  $^{14}$C and $^{15}$C  \cite{Cappuzzello_2015,Cappuzzello_epj,Bonaccorso}   have identified bumps  in the continuum, which have been proposed to be  a signature of the GPV. 

  Our studies have been based on the selection of a family of Woods-Saxon potentials \cite{10Li} which, after the inclusion of PVC effects, lead to  a  rather accurate reproduction of  the experimental energies of the low-lying levels in $^{10}$Li,  $^{11}$Be, $^{12}$B and $^{13}$C.  The single-particle wavefunctions have been determined solving the Schr\"odinger equation using an $r-$dependent effective mass $m^*(r)$ parameterized as a Fermi function,  going from $m^*= 0.72 m $ in the interior of the nucleus to the  value of the reduced mass of the system at large radii, simulating the typical behaviour of the mass resulting from Hartree-Fock calculations with Skyrme forces like SGII. 
The introduction of such $r-$dependent mass is not an essential feature of  our approach; similar results  can be obtained from parametrizations obtained fitting potentials  with a constant mass.

\begin{table}[h!]
\begin{center}
\begin{tabular}{|c|c|c| c |  c | c |  c |  }
\hline
&$\hbar \omega_{2^+}$ (MeV) & $\beta_2^+$ &$ V_{WS}$ (MeV) & $V_{ls} $ (MeV)
& $a_{WS} $ (fm) &  $R_{WS}$  (fm)  \\ \hline
$^{10}$Li &3.37& 0.68 & 64 & 14 & 0.75& 2.10 \\ \hline
$^{11}$Be&3.37& 0.71 & 72   & 18  & 0.72 & 2.14 \\ \hline
$^{12}$B& 3.80 &  0.57 & 77&  22& 0.78  &2.18 \\ \hline
 $^ {13}$C & 4.4 & 0.46  & 82&  27 & 0.73 & 2.23     \\ \hline
 \end{tabular}
\end{center}
\caption{\footnotesize Energies and deformation parameters of the low-lying $2^+$ state used in our calculations, together with the depths, the diffusivity and the radius  of the bare potentials parameterized 
according to Eq. (2.180) of ref. \cite{BohrI}.}
\label{bare_para}
\end{table}

 The energies and deformation parameters of the low-lying $2^+$ state used to calculate the PVC are reported in Table \ref{bare_para}.
The values adopted for $^{11}$Be and
$^{13}$C lie close to the experimental values of the corresponding $^{10}$Be
and $^{12}$C cores.
The values of $\hbar \omega_{2^+}$ and $\beta_{2^+}$ in
the odd-odd isotones $^{10}$Li and $^{12}$B have instead been fitted together with the Woods-Saxon parameters,
so as to optimize the properties of the calculated energy spectrum as compared to experimental data.
These quantities should be interpreted as a property of the even-even core in the presence of an extra proton. 

\begin{figure}[h!]
    \centering
     \includegraphics[width=0.8\linewidth]{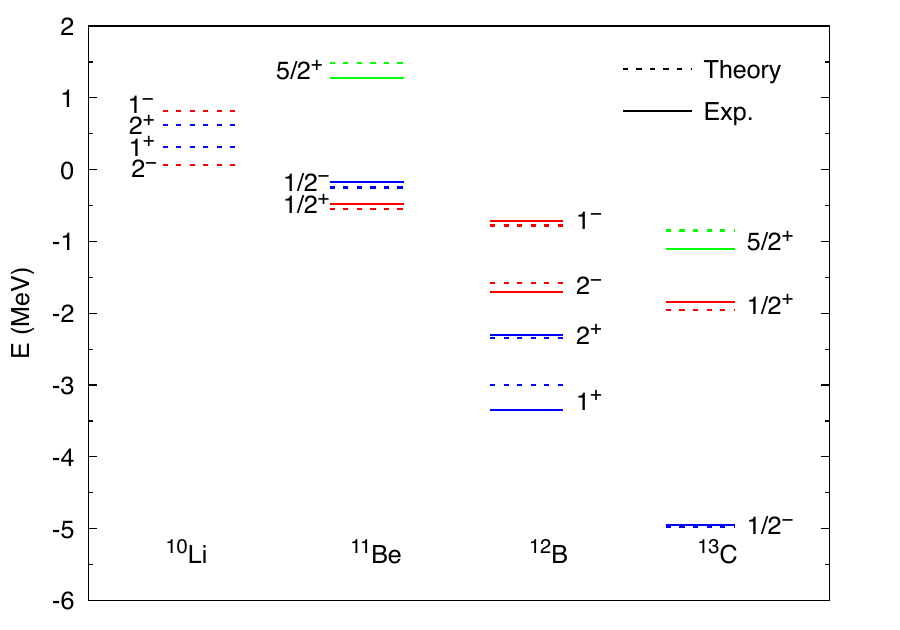}
    \caption{\footnotesize The experimental energies of the low-lying states in the N = 7 isotones
$^{10}$Li, $^{11}$Be, $^{12}$B and $^{13}$C are shown by solid lines. The corresponding theoretical energies are displayed by dashed lines. States based
on $1/2^-, 1/2^+$ and $5/2^+$ neutron configurations are shown by blue, red, and green 
lines, respectively. From \cite{10Li}.}
    \label{fig:compare_levels}
\end{figure}

The values of the Woods-Saxon parameters are also listed in Table  \ref{bare_para}.
The radius, depth and spin-orbit strength of the potential evolve smoothly as a function of $N$, while the  diffusivity remains essentially constant.
The comparison of the resulting theoretical spectra with the  experimental energies is shown in Fig. \ref{fig:compare_levels}.

\subsection{$^{11}$Be}

\begin{figure}[b!]
    \centering
    \includegraphics[width=0.7\linewidth]{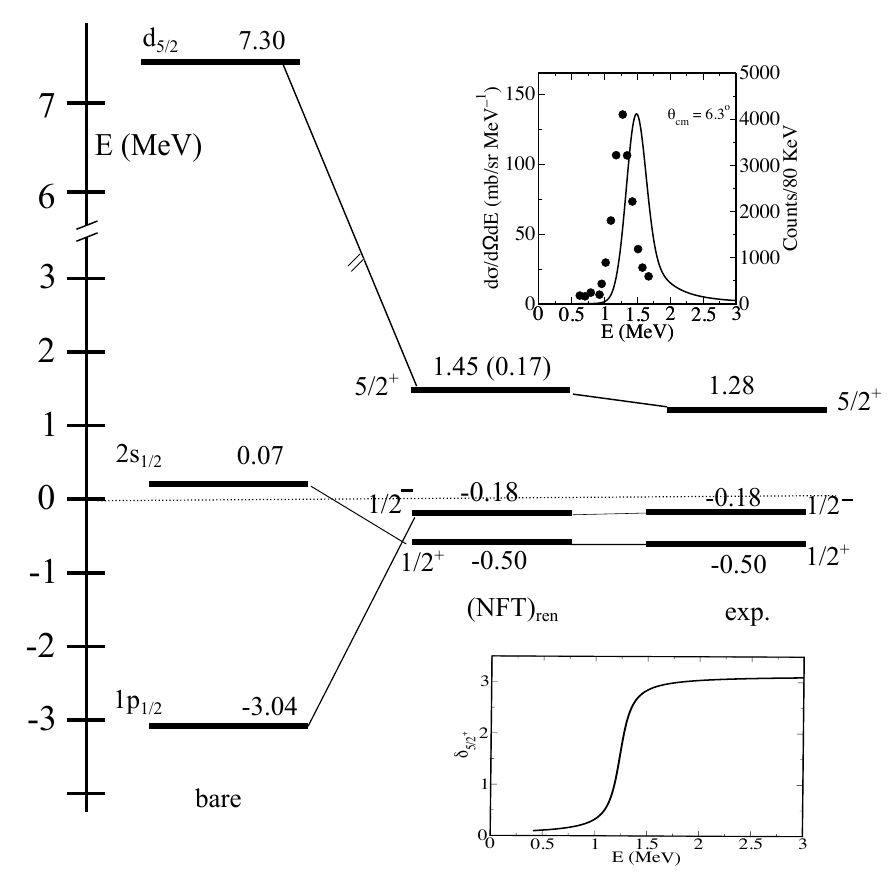}
\caption{\footnotesize  Low-lying spectrum of $^{11}$Be. (bare): unperturbed single-particle levels;
solution of the bare mean field.
((NFT)$_{ren})$: dressed levels; 
(exp.):  experimental values.
The number on each thick horizontal line 
gives  the energy of the state in MeV.
The calculated strength function of the $\widetilde{5/2^+}$ resonance is shown in the inset on the upper right hand side,
compared with data \cite{Schmitt}. The associated  elastic phase shifts are shown in the inset on the lower right hand side \cite{11Be}. }    
    \label{fig:levels_11Be}
\end{figure}

The importance of core admixture  in halo nuclei was clearly demonstrated by the analysis of the  $^{11}$Be(p,d)$^{10}$Be transfer reaction performed at GANIL laboratories in inverse kinematics \cite{Fortier,Winfield}. 
The most significant finding was the precise measurement of the cross section populating  the first excited $2^+$ state of $^{10}$Be (located at 3.37 MeV). The population of this state demonstrated that the ground state
of $^{11}$Be is not a simple neutron moving around an inert sphere, but contains a 0.16 $[2+ \otimes 1d]$ core-excitation admixture.
This  result suggested that the core of a halo nucleus is highly dynamical and deformable, validating particle-vibration coupling models like the one proposed in this contribution,  in the spirit of
the unified model of Bohr and Mottelson.


Turning now to our results,
in Fig. \ref{fig:levels_11Be}  the energies  of the single-particle levels calculated in $^{11}$Be with the 'bare' Woods-Saxon potential are compared with the energies obtained after the inclusion of the self-energy diagrams. The distance between $s_{1/2}$ and $p_{1/2}$ waves in the bare mean field potential  is about 3 MeV in the case of $^{11}$Be. Self-energy effects then lower the energy of the $1/2^+$ orbital by about 500 keV and increase the energy of the $1/2^-$ orbital by 2.9 MeV, leading to parity inversion.
There is no low-lying $d_{5/2}$ level in the bare mean field, and the renormalized  $\widetilde{5/2^+}$ state  is a resonance with a strong many-body character.

\begin{figure}[b!]
    \centering
    \includegraphics[width=.9\linewidth]{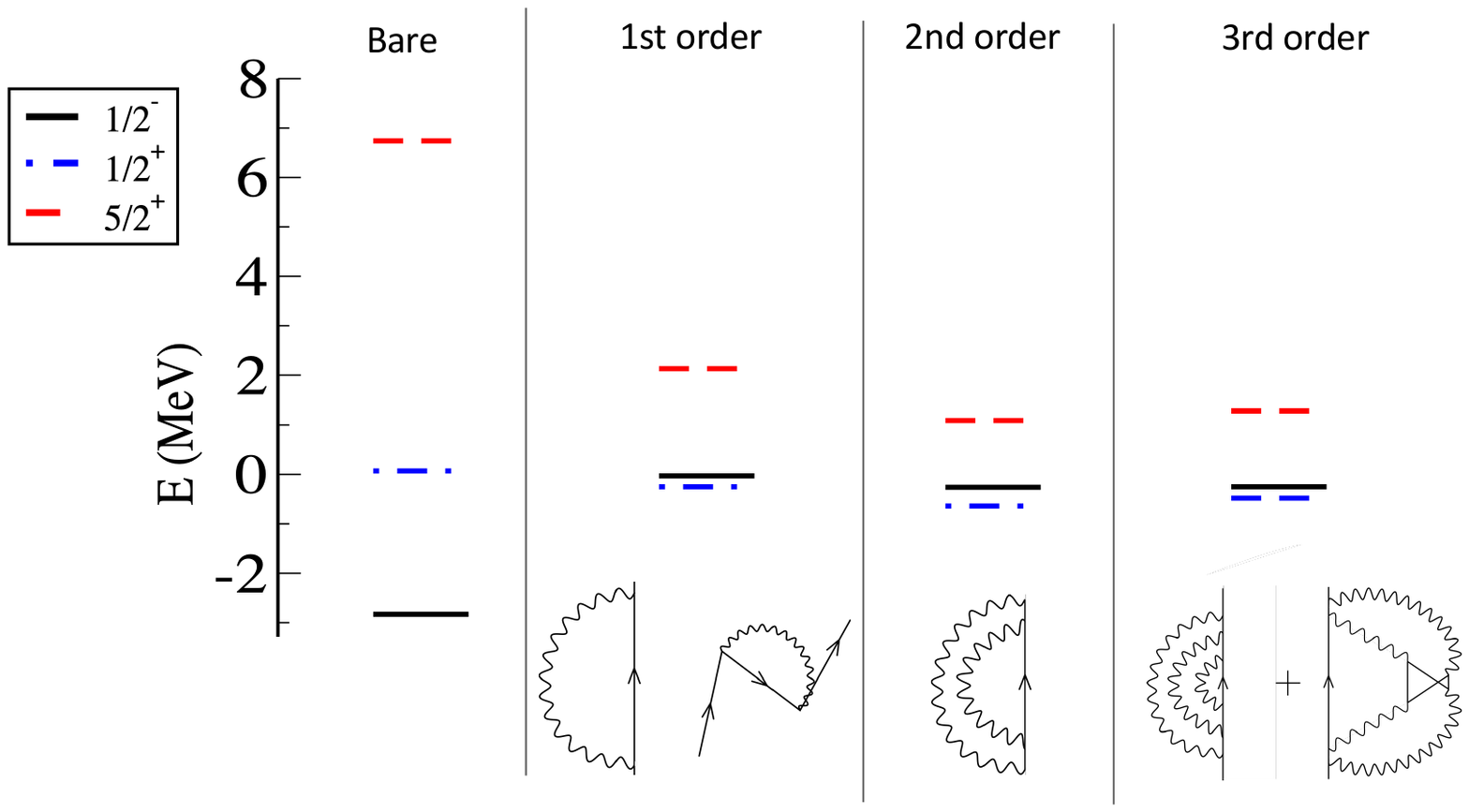}
    \caption{\footnotesize Theoretical spectrum of low-lying states of $^{11}$Be, calculated in different perturbation orders  in NFT. The lowest
    $1/2^-$,$1/2^+$ and $5/2^+$ levels are shown by solid, dot-dashed 
    and dashed lines respectively. In the case of $5/2^+$, the energy of the resonant state is indicated. Representative diagrams included at each order (see Fig. 2 and 3) are drawn at the bottom of the figure. From \cite{OurVenice}.} 
    \label{fig:ConvergenceVenezia}
\end{figure}
We have studied in detail the convergence of the calculation when more than one-phonon configurations, as well as the phonon-phonon interaction,
discussed in previous sections, 
are included in  the 
self-energy (see Fig.\ref{fig:ConvergenceVenezia}). It can be appreciated how the results stabilize already at third order. The second order rainbow diagram acts in an important way on the 
$\widetilde{5/2^+}$ state, while parity inversion already appears at first order. At third order the rainbow diagram and the "butterfly" correction essentially compensate each other. Furthermore, we have checked that the rainbow series for orders higher than three gives a negligible contribution. 

In our approach the energies of the dressed states do not represent actual predictions, because  the experimental values have been used to fit the
mean field.
The wavefunctions of the dressed states, including components up to one phonon, can be schematically written as 
\begin{align}
  \ |\widetilde{1/2^+} \rangle  &=  \sqrt{0.80} |s_{1/2}\rangle + \sqrt{0.20} |(d_{5/2}\otimes 2^+)_{1/2^+}\rangle    \\
  |\widetilde{1/2^-}\rangle &=  \sqrt{0.84} |p_{1/2}\rangle + \sqrt{0.16}  |((p_{1/2},p^{-1}_{3/2})_{2^+}
\otimes 2^+)_{0+}, p_{1/2}\rangle \\
|\widetilde{5/2^+}\rangle & =\sqrt{0.49} |d_{5/2}\rangle+ \sqrt{0.23}|(s_{1/2}\otimes 2^+)_{5/2^+} \rangle  \nonumber  \\
& + \hspace{0.5cm}\sqrt{0.28} |(d_{5/2}\otimes 2^+)_{5/2^+} \rangle. 
\label{be_formfactors}
\end{align}  

Folding the theoretical calculation with a gaussian curve of FWHM = 220 keV, representing the experimental resolution, estimated from the width of the peaks of the discrete states, we obtain the line shape of the dressed $\widetilde{5/2^+}$ resonance shown in the inset of Fig. \ref{fig:levels_11Be}. It exhibits  a FWHM of 400 keV, in good agreement with the experimental results from the $^{10}$Be(d,p)$^{11}$Be(5/2$^+$) one-nucleon transfer reaction \cite{Schmitt}.  It is remarkable that such a width is considerably larger than the  width equal to approximately 100 keV associated with the $^9$Be(t,p)$^{11}$Be two-nucleon transfer reaction \cite{Kelley}, and also larger than that exhibited by the calculated elastic phase shifts (170 keV, see again Fig. \ref{fig:levels_11Be}).

The quadrupole admixture in the ground state leads to corrections in the mean square radii and in the B(E1) transition strength. The resulting values are  $<r^2>^{1/2}$ = 2.48 fm, and  B(E1) ($1/2^-$ $\to$ $1/2^+$) = 0.11 ${\rm e^2 fm^2}$, to be compared with the experimental values 2.466 $\pm$ 0.015 fm \cite{Krieger} and 0.102 $\pm$ 0.002 \cite{Kwan}. 

\begin{figure}
    \centering
    \includegraphics[width=0.45\linewidth]{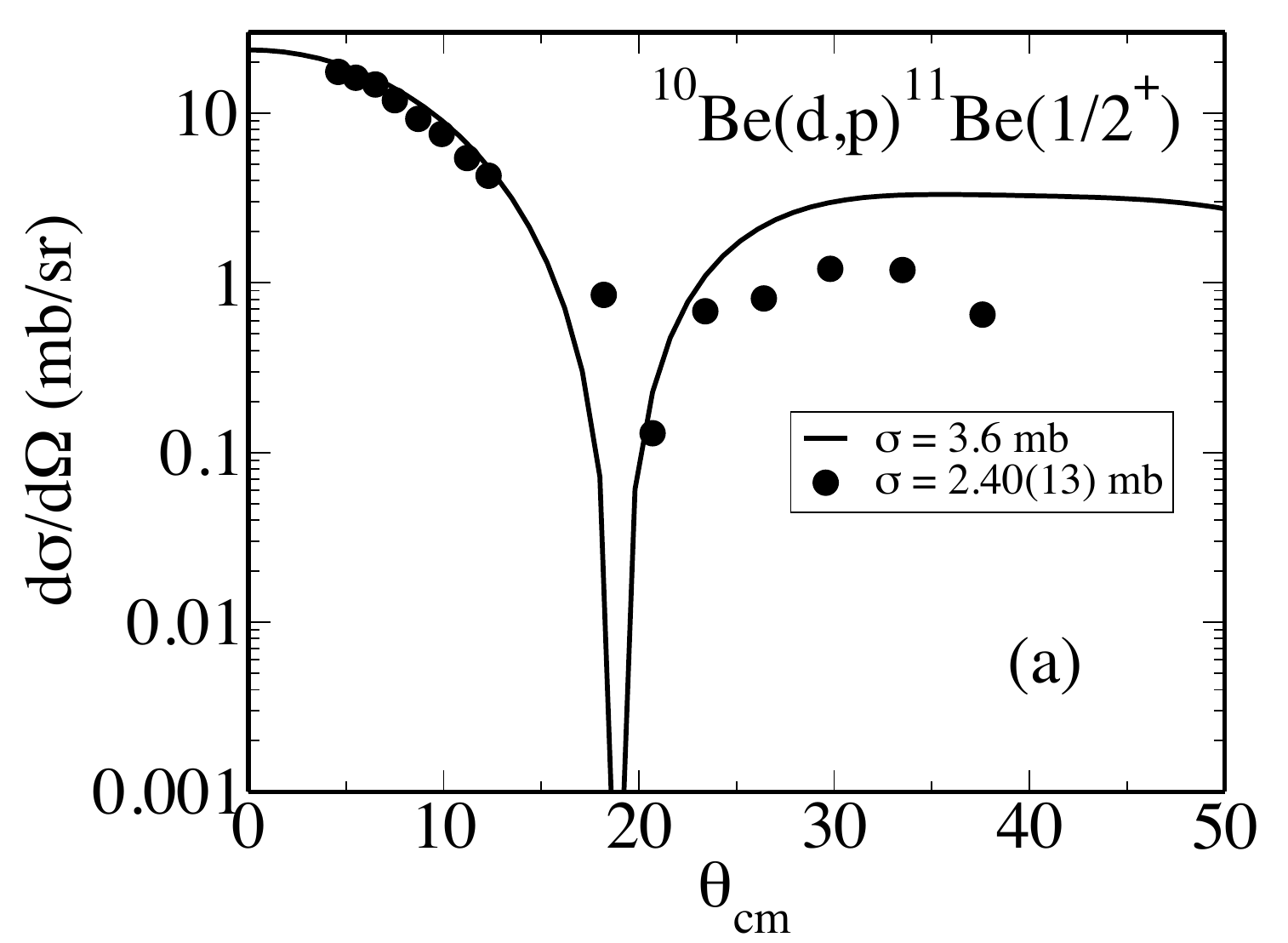}
     \includegraphics[width=0.45\linewidth]{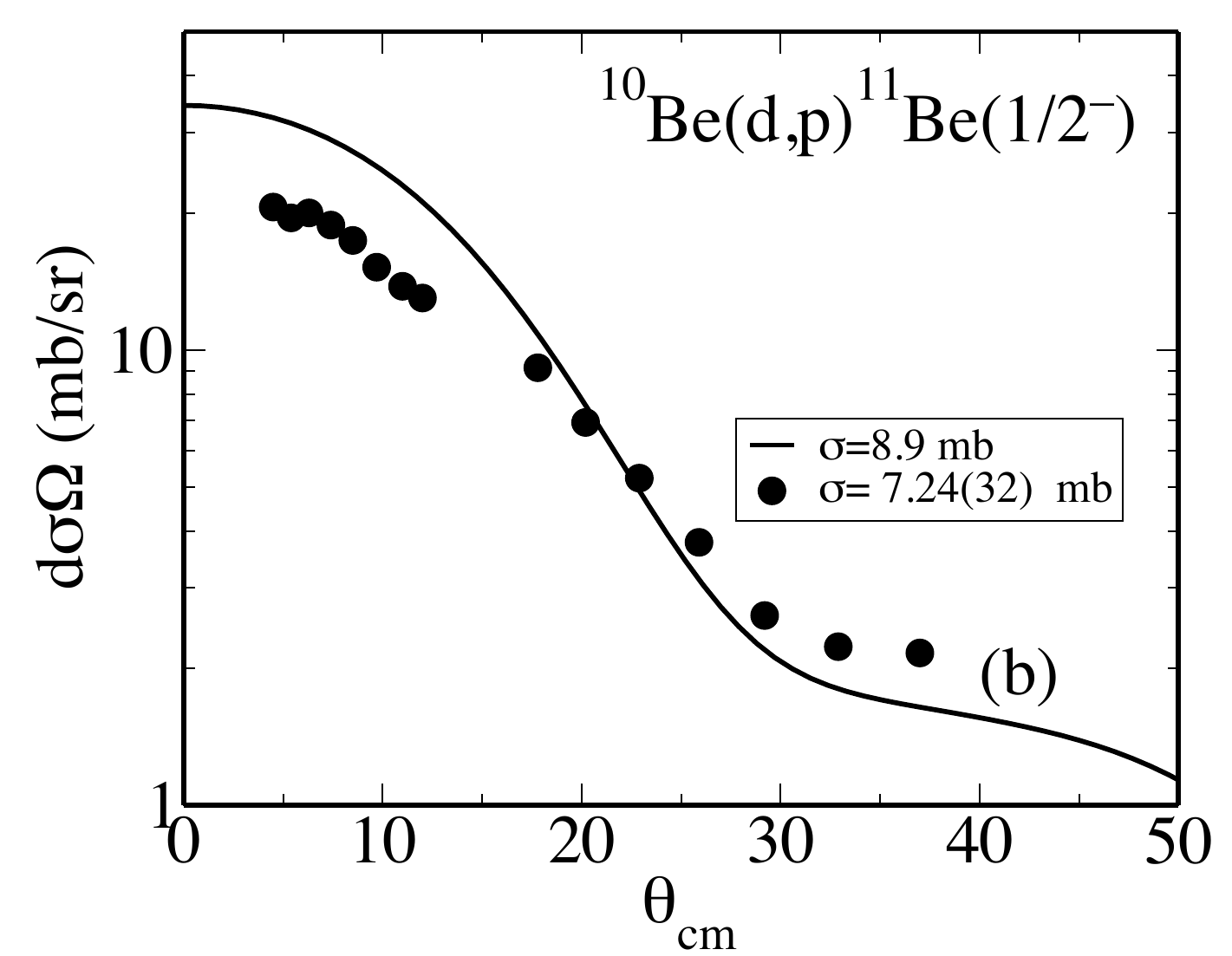}
            \includegraphics[width=0.45\linewidth]{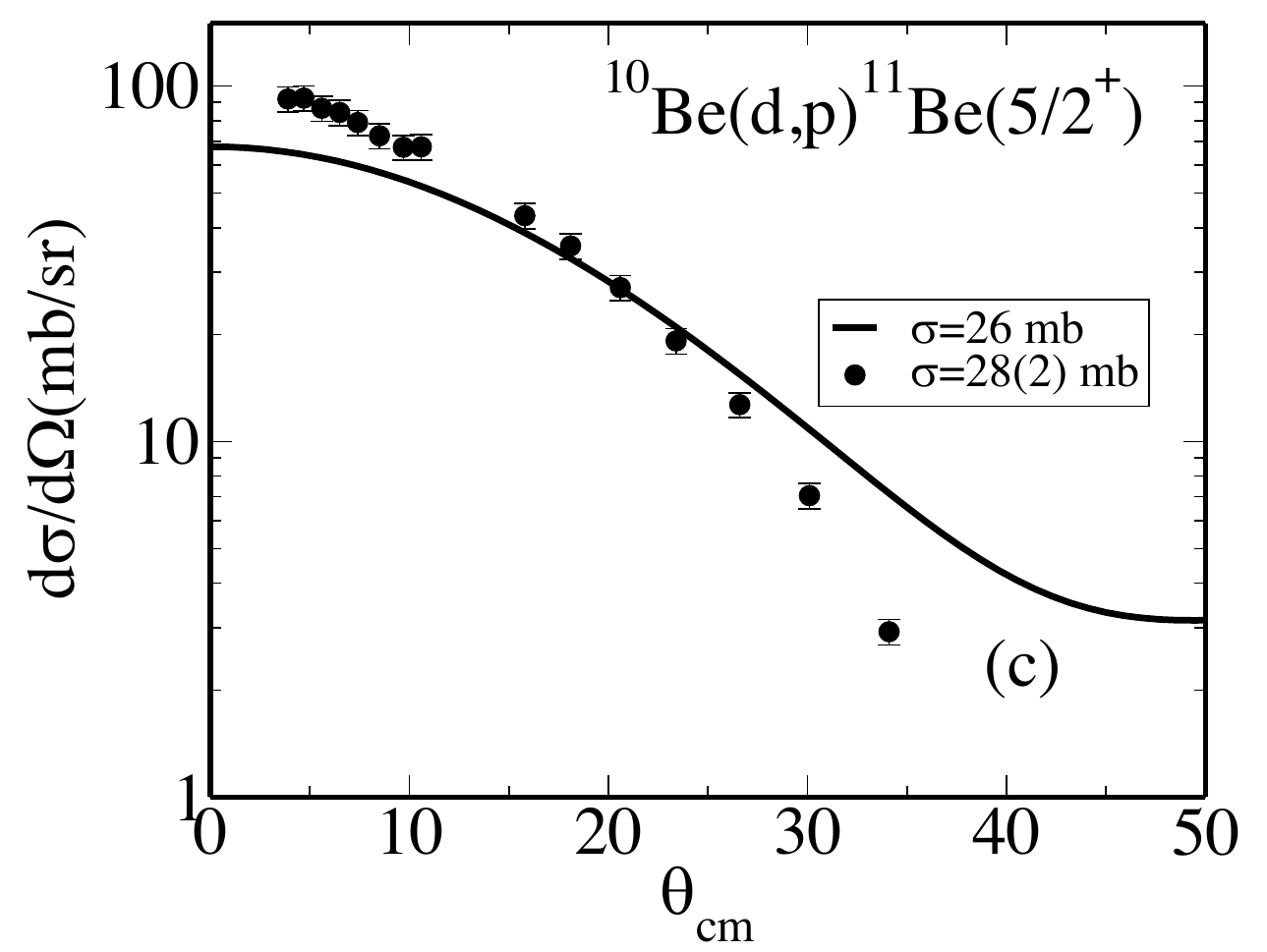}
              \includegraphics[width=0.45\linewidth]{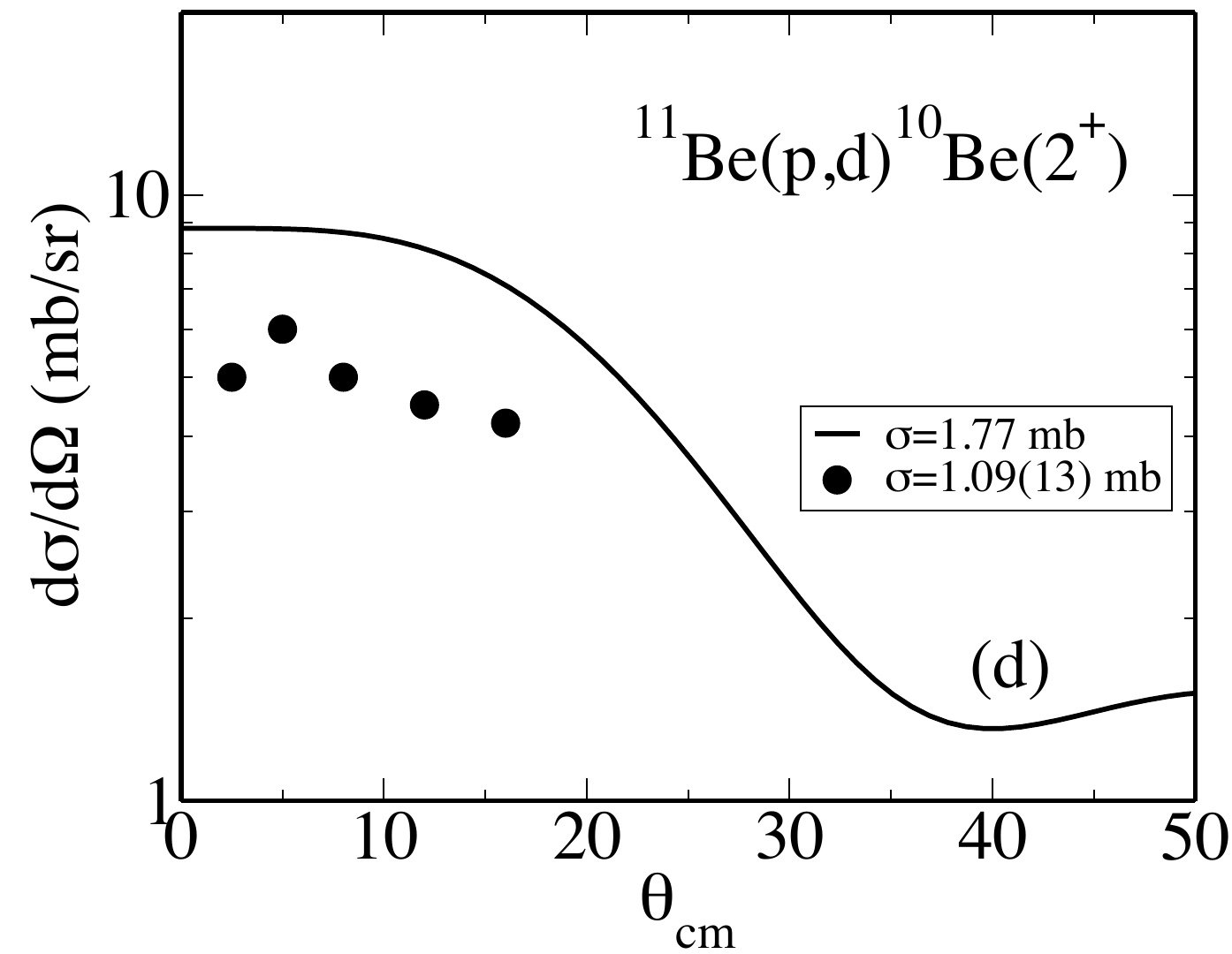}
\caption{\footnotesize (a-c) Absolute differential and (insets) summed cross sections associated with the reactions $^2$H($^{10}$Be,$^{11}$Be)$^1$H at E=107 MeV, populating the $1/2^+$ , $1/2^-$ and $5/2^+$ states, compared to experimental data \cite{Schmitt}. (d) The same as before, but for the reaction $^1$H($^{11}$Be,$^{10}$Be)$^2$H at E=388.3 MeV, populating the $2^+$ state \cite{Winfield}. These calculations, as well as the cross section shown in the inset of Fig. \ref{fig:levels_11Be}, were performed making use of global optical potential parametrizations for the proton \cite{KoningDelaroche2003} and deuteron \cite{hanDeuteronGlobalOptical2006} channels. From \cite{11Be}.}
    \label{fig:11Be_cross}
\end{figure}

The wavefunctions (15-17) can be used as form factors  for the calculation of the cross sections  of one-nucleon transfer reactions $^{10}$Be(d,p) and $^{11}$Be(p,d),  that represent tests of our theory.  The 
$^{11}$Be(p,d) reaction populating the $2^+$ state of the $^{10}$Be core is of particular interest, because it probes the quadrupole admixture in the ground state, which represents the cornerstone of our approach. Theoretical cross sections calculated in the Distorted Wave Born Approximation (DWBA) are compared to experimental data in Fig. \ref{fig:11Be_cross}.  These calculations rely on optical potentials taken from systematics.


\subsubsection{Halo nuclei: vibrational or deformed? }

An alternative and rather widespread approach to account for the deformability of the halo nuclei cores assumes a strong coupling between the valence particle and a  static deformed core, as in the Nilsson model. 




An interesting example is the work by  Hamamoto and Shimoura \cite{hamshim}  on the properties of $^{12}$Be and $^{11}$Be, and in particular on the measurements of $B(E2; 0_2^+ \to 2_1^+)$ and $B(E0; 0_2^+ \to 0_1^+)$ in $^{12}$Be. The authors use a deformed Woods-Saxon potential and obtain a rather good description of the single-particle motion and transition probabilities,  demonstrating that a simple model of neutrons in a deformed potential can explain a variety of spectroscopic data in $^{11,12}$Be. The large $B(E1)$ values are a direct consequence of the halo-like wave functions and the mixing of $s$ and $p$ orbits in the deformed field. The breakdown of the $N=8$ shell closure is seen as a transition to a regime where deformation and weak binding energy dictate the nuclear structure.

In a successive work based on this model,  spectroscopic factors for different (p,d) and (d,p) reactions involving $^{11}$Be have been obtained by a best fit analysis of the expansion coefficients of the Nilsson levels on a spherical basis  \cite{Macchia}.  The coefficients obtained in \cite{Macchia} are in good agreement with those calculated in \cite{hamshim} for the [2201/2] Nilsson orbital  and correspond to a deformation $\beta_2 \approx 0.8$, but are very different in the case of the  [1101/2] orbital. The amplitude of this orbital on the $p_{3/2}$ spherical state  is two times larger than that obtained in \cite{hamshim}, and would be associated with an unphysical, much larger deformation.

\begin{figure}[b!]
    \centering
    \includegraphics[width=0.8\linewidth]{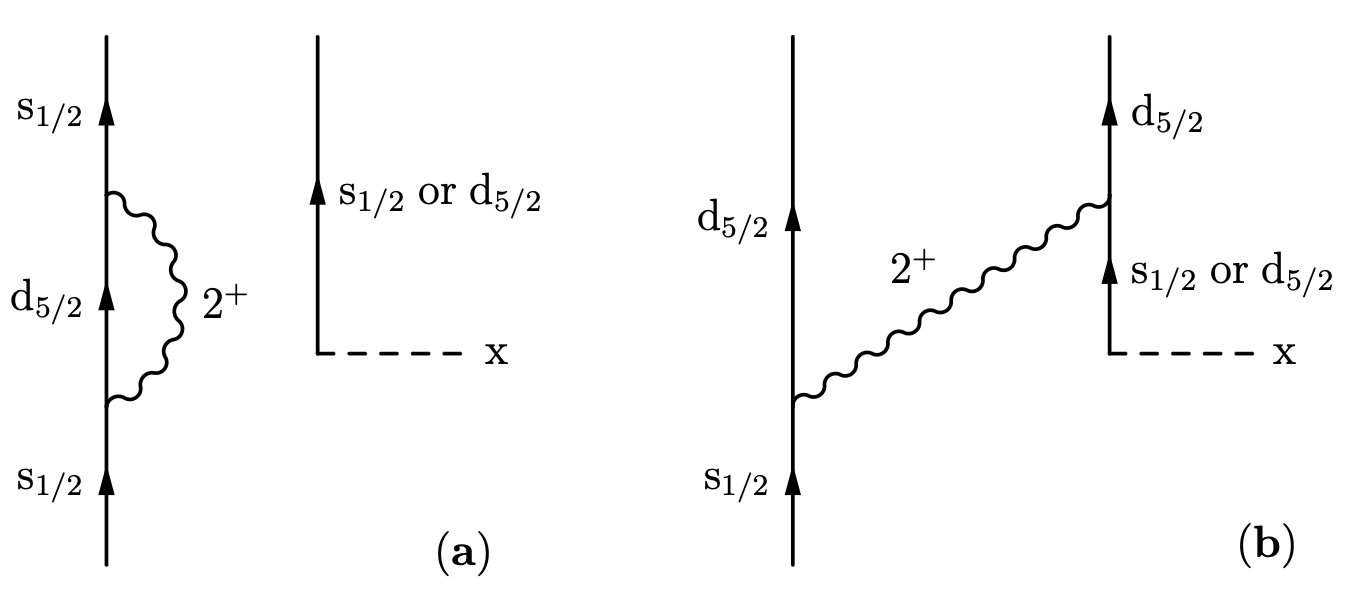}
\caption{\footnotesize Induced interaction process influencing the population of $(d_{5/2})^2$ configuration in $^{12}$Be by the  $^{11}$Be(d,p)$^{12}$Be reaction. The dashed line represents the action of  the transfer operator.
(a) The wavefunction of the  $1/2^+$ ground state in $^{11}$Be has a sizeable $[d_{5/2} \otimes 2^+]_{1/2^+}$ component. The transfer operator can populate
the empty $s_{1/2}$ or $d_{5/2}$ orbital. (b) The two $s_{1/2}$ orbitals, or the $s_{1/2}$ and the $d_{5/2}$ orbitals, can dynamically exchange  the $2^+$ phonon, and populate the final
$(d_{5/2})^2$ configuration.}
    \label{fig:Be_induced}
\end{figure}

Moreover, in the case  of the $^{11}$Be(d,p)$^{12}$Be reaction, populating different states ($0^+_1, 0^+_2 $ and $ 2^+_1$)   in $^{12}$Be, the  best fit is not able to reproduce the experimental spectroscopic factors. 
The dynamics of the core surface becomes crucial in $^{12}$Be in which the induced interaction, resulting from   the phonon exchange between the halo neutrons may play an important role. An example related to the population of $(d_{5/2})^2$ configurations is shown in Fig. \ref{fig:Be_induced}.  The $^{11}$Be ground state contains an admixture of the $d_{5/2} \otimes 2^+$ configuration, as discussed above and shown in Fig. \ref{fig:Be_induced}(a). If the transfer operator creates a particle on the $s_{1/2}$ or $d_{5/2}$ orbital, it becomes possible to reach the $(d_{5/2})^2$ configuration through the exchange of the virtual $2^+$ phonon, as depicted in  Fig. \ref{fig:Be_induced}(b).
 This process  cannot be taken into account in an approach based on a  static deformed mean field. As stated by Hamamoto and Shimoura \cite{hamshim}, calculated spectroscopic factors are likely to be affected by the fact that "the inactive core nucleus $^{10}$Be in our present model is different from the observed nucleus $^{10}$Be."


A constant deformation of the core is also assumed in more dynamical studies based on 
the particle-rotor model (PRM), which  
take into account  the existence of different  rotational states and their coupling to the motion of the valence neutron.
 In an early study, \cite{Nunes} a significant admixture of the $2^+$ core rotational state was found for the   $\widetilde{1/2^+}$ and $\widetilde{5/2^+}$ states in $^{11}$Be, similarly to what was found in \cite{Winfield,11Be}. The energy of  the  $\widetilde{1/2^+}$ state was lowered with respect to the bare  $2s_{1/2}$ orbital. However, the shift was not sufficiently large  
 and it was necessary to assume a very large spin-orbit splitting to reproduce  the parity inversion of the ground state. In other works, a parity-dependent mean field was introduced instead.

 \begin{figure}[h!]
    \centering
    \includegraphics[width=0.9\linewidth]{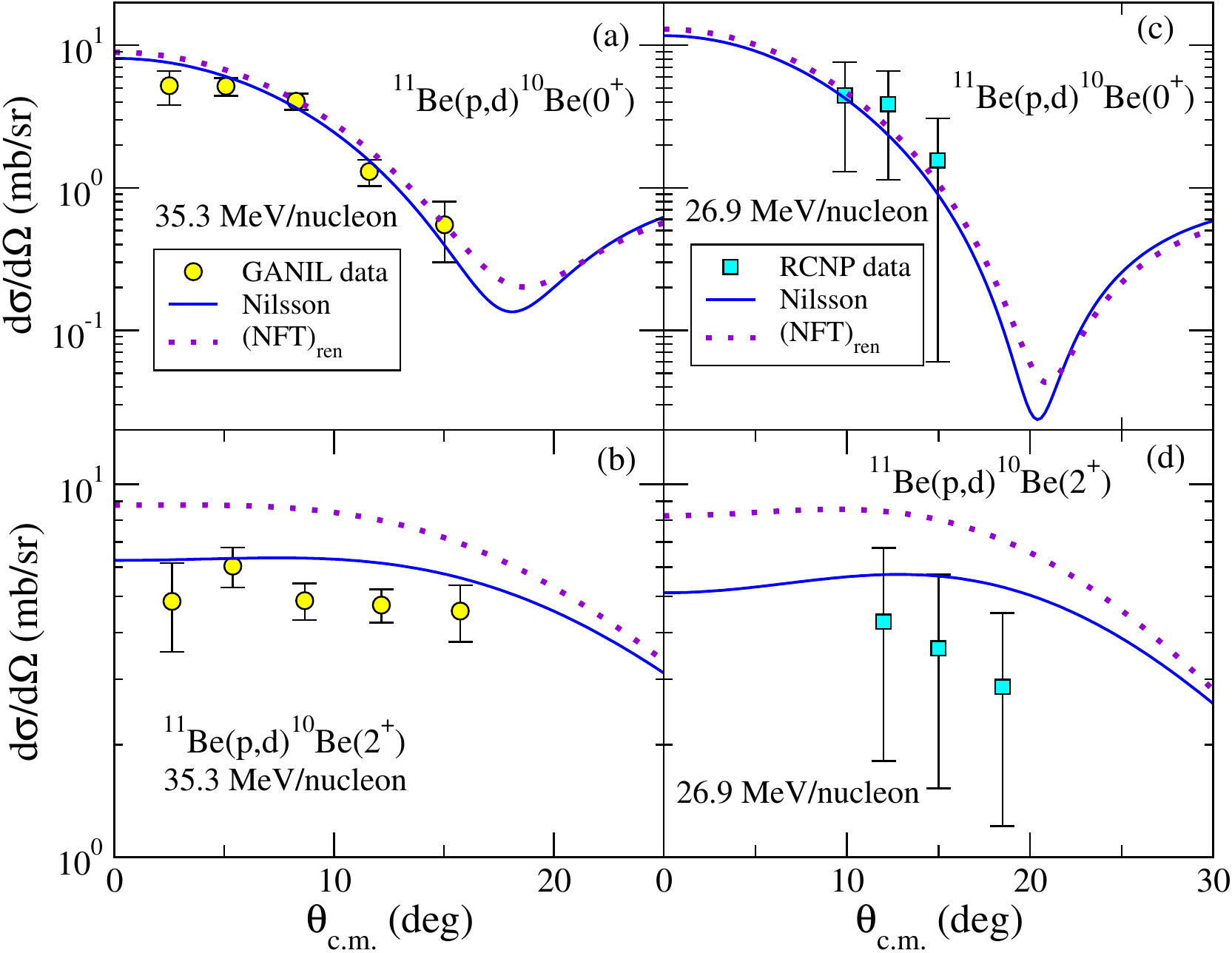}
    \caption{\footnotesize Differential cross section for the $^{11}$Be(p,d) transfer reaction populating the ground state  (upper panels) and
    the lowest $2^+$ state (lower panels) in $^{10}$Be. 
    In panels (a) and
(b), the calculations carried out according to  the PRM (Nilsson) \cite{punta2023} and to NFT for 35.3 MeV/nucleon are compared with the experimental data from GANIL \cite{Fortier,Winfield}. The calculations for 26.9 MeV/nucleon are compared with the experimental data from RCNP \cite{Jiang} in panels (c) and (d). The figure is due to Pedro Punta de la Herr\'an \cite{Punta_phd}.}
    \label{fig:punta}
\end{figure}

\begin{figure}
    \centering
\includegraphics[width=0.5\linewidth]{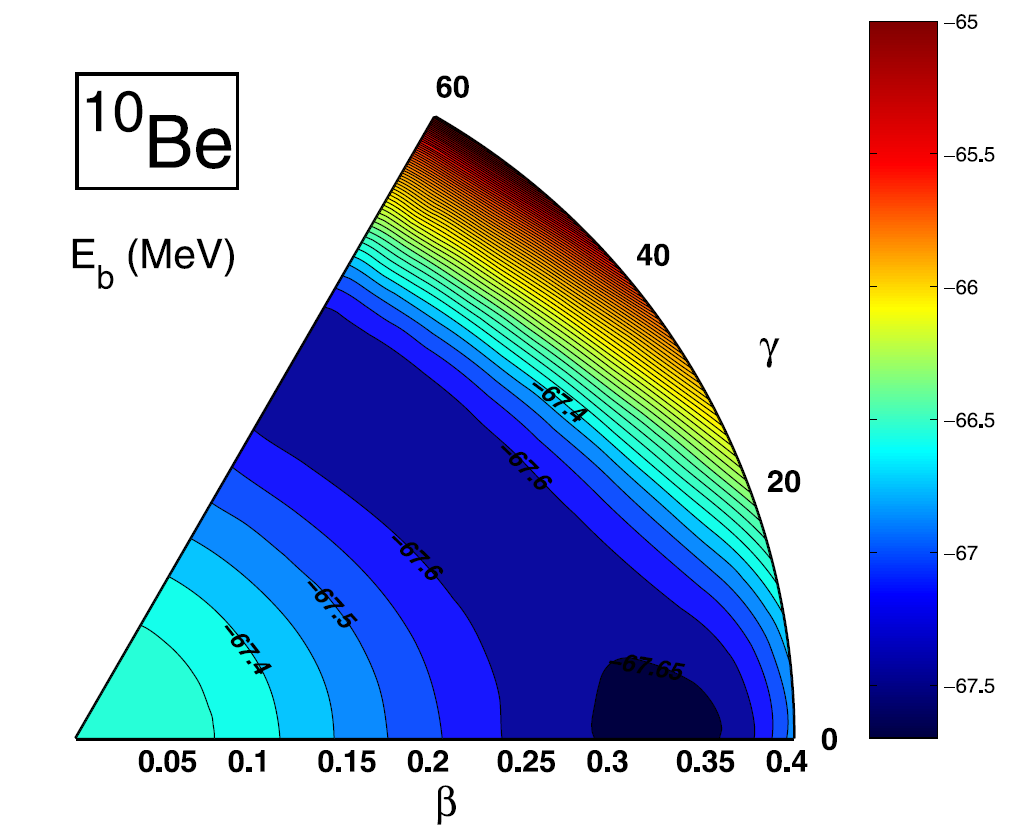}
\includegraphics[width=0.8\linewidth,trim=4pt 120pt 1pt 4pt,clip]{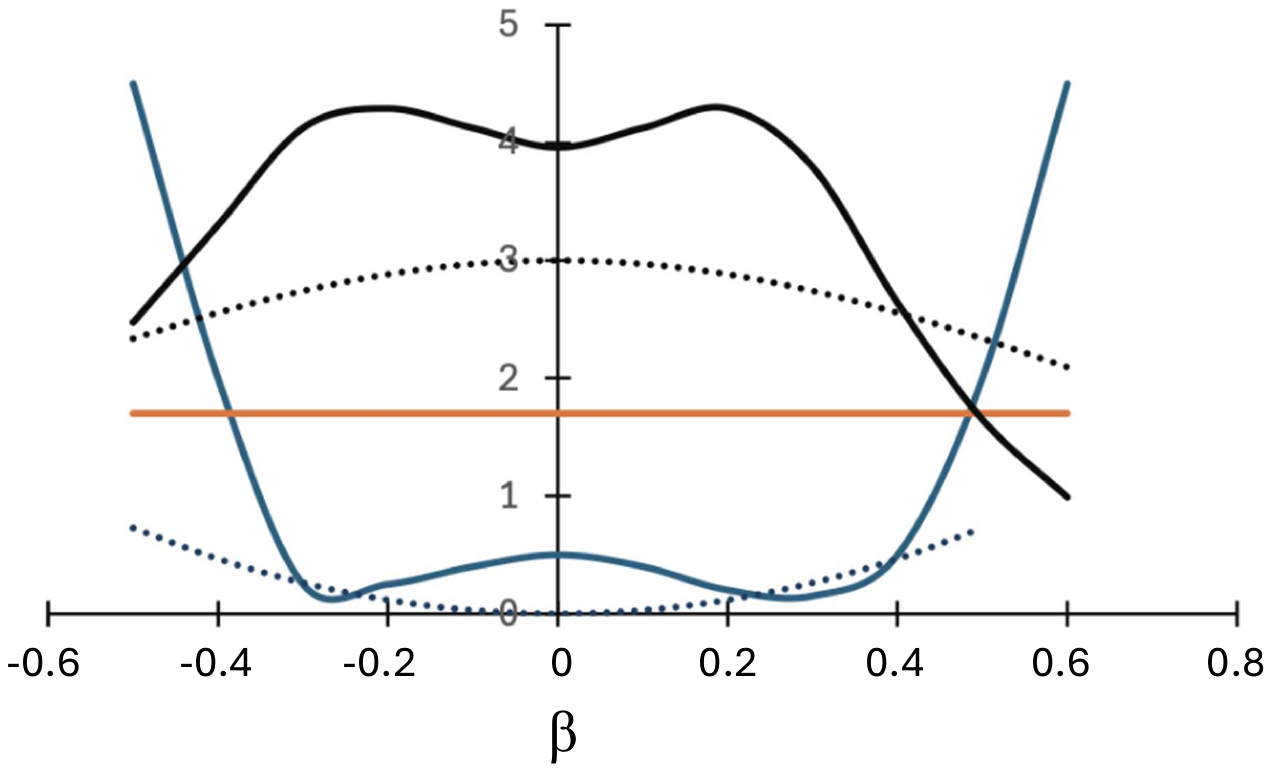}

    \caption{\footnotesize Top panel: energy surface of $^{10}$Be, taken from \cite{zhang}. Bottom panel: the solid blue line represents the  energy surface of $^{10}$Be  along the $\gamma = 0^o$ ($\beta>0) $ and along the $\gamma= 60^o$ ($\beta <0)$ axis extracted from the top panel. The solid black line displays the corresponding ground state wavefunction, assuming a constant mass parameter.   The orange horizontal line represents the zero-point energy $1/2 \; \hbar \omega_2$, using the experimental value $\hbar \omega_2$ = 3.4 MeV.  The lower dotted line  shows the harmonic oscillator  potential energy obtained from  $\hbar \omega_2$ and the experimental $B(E2)$ value,
   while  the upper dashed line shows the  corresponding wavefunction.}
    \label{fig:paco_10be}
\end{figure}

 The PRM matrix element connecting a single-particle state $lj$ with a particle-rotor configuration $ ((l'j') \otimes R)_{lj}$ is, apart from a small reorientation term, equal to that of the PVC scheme 
 (see Eq.(5A-5) and Eq. (6-209)  in \cite{BohrII}), provided that the root mean square value of the quadrupole deformation in the vibrational picture is equal to  the  fixed deformation of the rotor. 
 This explains the similarity of the results found for the $\widetilde{1/2^+}$ state.
More advanced implementations of the PRM have been used to calculate form factors for successful calculations of  cross sections of breakup \cite{dediego} and  one-nucleon transfer reactions \cite{punta2023}, including reaction populating the $2^+$ excited state of the core, demonstrating the importance of core excitation. An example is shown in Fig. \ref{fig:punta}, where the  cross sections of the (d,p) reaction
on $^{11}$Be populating the ground state and the $2^+$ state in $^{10}$Be 
calculated with NFT (already shown in Fig. \ref{fig:11Be_cross}(d)) and  PRM are compared. They are in good agreement between themselves and with the data for the 
ground state, while they overestimate the cross sections for the $2^+$ state, 
which are slightly better reproduced by the PRM.

 However, the implementation of  Pauli principle in PRM calculations is problematic, due to the transformation from the intrinsic to the laboratory frame \cite{Urata,punta2023,Punta2,Watanabe}. The  coupling to $h \otimes b$  configurations discussed above and related to the self-energy correlation diagram   associated with core shape fluctuations is absent in the PRM, which postulates an inalterable rigid rotor as core. As a consequence, the energy of the $p_{1/2}$ level is not increased and the process is mocked up by introducing a strong  spin-orbit splitting or a parity-dependent mean field.  
 Furthermore, when more complex configurations (i.e. containing more than one phonon in PVC or higher rotational states in PRM) are needed, as it happens with the $\widetilde{5/2^+}$ state in $^{11}$Be, the two schemes show differences. 
 

The use of static deformed WS potentials, as outlined above, may represent a too simplified picture for the nuclei we are considering. 
Some further insight on this matter may be obtained by noting that the energy surfaces in the quadrupole deformation space for these nuclei are typically very shallow.
An example is shown in Fig. \ref{fig:paco_10be} (top), taken from \cite{zhang}. More recent calculations share the same qualitative features \cite{Geng,Shen}.  One can observe an absolute  minimum at $\beta_2 \approx 0.35^o, \gamma=0^o$, suggesting a prolate nucleus. However, such a minimum is quite shallow.  In Fig. \ref{fig:paco_10be} we sketch the potential energy along the $\gamma=0^o$ and $\gamma=60^o$ axes, evidencing that the depth of the minimum is about 0.5 MeV. We notice that the energy  of the first $2^+$ state is $\hbar \omega_2 = 3.4$ MeV.  
The corresponding zero-point energy  1/2 $\hbar \omega_2$  is also plotted in  Fig. \ref{fig:paco_10be} (bottom), together
with the associated ground state wavefunction, which samples a large interval in $\beta_2$, including positive and negative values.  The absolute minima of the energy surface, located near $\beta = \pm 0.3$ are  therefore of little significance. 
The first $2^+$ excited state would  lie 3.4 MeV higher and would sample an even broader range of deformations, which would be hardly compatible with the image of a rigid rotor having  a fixed static deformation. 
It can also be seen  that the potential energy  and the wavefunction are in qualitative agreement with those obtained from the experimental values of $B(E2)$ and $\hbar \omega_2$ in the harmonic oscillator approximation (see dashed curves).  Nonetheless, it is evident that a pure harmonic picture is just a first approximation; the energy surface indeed shows a clear departure from a parabolic shape. The anharmonicities may be studied by considering phonon-phonon interactions, as that represented by the "butterfly" NFT diagram mentioned above in the case of the $5/2^+$ level in $^{11}$Be (see Figs. \ref{fig:farfalla} and \ref{fig:ConvergenceVenezia}), where it leads to a good agreement with the data.
 This diagram will induce  a splitting of the two-phonon triplet $[2^+ \otimes 2^+]_{0^+,2^+,4^+}$ in $^{10}$Be,  possibly reproducing  the $4^+$ state energy and electromagnetic transitions in $^{10}$Be without invoking the rotational picture \cite{vonoertzen}. This work is in progress.
It is useful also to  refer to a study of the wavefunction of  $^{10}$Be obtained by the Generator Coordinate Method, sampling an energy surface in the 
${\beta,\gamma}$ plane, which displays the same  qualitative features described above, generated by a microscopic  Antisymmetrized Molecular Dynamics calculation \cite{Kanada}.

In conclusion, a dynamical description that takes into account the large surface fluctuations, seems more suitable to capture the main features of these light systems.

\subsection{$^{10}$Li}

In the case of the nucleus $^{10}$Li, we have calculated  the 
 self-energy 
matrix  $\Sigma(k,k',\omega)$ for the single particle levels $k,k'$ up to 50 MeV, in a box of radius 60 fm.
We have considered the coupling between the valence neutron and the $^{9}$Li core. The latter is a $^8$He nucleus modified by the presence of a proton on the $p_{3/2}$ orbital,  and we have determined its energy and deformation parameter by a best fit analysis. The resulting values 
($\hbar\omega_2=3.37$ MeV, $\beta_2=0.68$) nicely 
interpolate the corresponding values  in $^{8}$He and $^{10}$Be.

In the following we will present the results obtained for the renormalized $\widetilde{ 1/2^+}$ and  $\widetilde {1/2^-}$ waves, which are virtual and resonant states lying in the continuum and  essentially 
 determine the  spectrum up to about 2 MeV. 
The $\widetilde{1/2^+}$  scattering length  is $\alpha=$ - lim$_{k\to 0} \; {\rm tan}(\delta_{1/2^+})/k = - 8$ fm,
	 corresponding to the energy   \cite{Landau:81,Friedrich:16} $\epsilon_{\widetilde{1/2}^+} = 
	\frac{\hbar^2\kappa^2}{2m} 	=  0.32 \; {\rm MeV} $,
	 where $ \kappa = 1/\alpha$. 
The eigenfunction of a state  lying close to $\epsilon_{\widetilde{1/2}^+} $
and thus representative of this virtual state is,
		\begin{equation}
		\label{eq1}
	 \ket{\widetilde  {1/2}^+}  =\sqrt{0.98}\ket{s_{1/2}}   
	 +\sqrt{0.02}\ket{\left(d_{5/2}\otimes 2^+\right)_{1/2^+}}.
	 \end{equation}
Similarly, the resonant  $\widetilde{1/2^-}$ state can be written as 
	\begin{align}\label{eq2}
	& \ket{\widetilde{1/2}^-}   = \nonumber  \\  & \sqrt{0.94}\ket{p_{1/2}}  
 +\sqrt{0.07} \;  \;\ket{((p_{1/2},p_{3/2}^{-1})_{2^+}\otimes 2^+)_{0^+},p_{1/2};1/2^-}.
 \end{align}
 The peak and the width of the resonance are $\epsilon_{\widetilde{1/2}^-} = $ 0.50 MeV, and 
$\Gamma_{\widetilde{1/2^-}}  =  2\left(\frac{d\delta_{\widetilde{1/2^-}}}{dE}|_{\epsilon_{\widetilde{1/2^-}}}\right)^{-1}= 0.35 \; {\rm MeV}$ respectively.
		The parallel between these results and those previously shown  for $^{11}$Be is apparent. The dressed  neutron  
	couples to the $1p_{3/2}^{-1}(\pi)$ proton hole  leading
	to the doublets ($1^-,2^-$) (${\widetilde{1/2}^+}  \otimes  p_{3/2}^{-1}(\pi)$) and ($1^+$, $2^+$) (${\widetilde{1/2}^-}  \otimes p_{3/2}^{-1}(\pi)$). 
      In keeping with  a well established approach \cite{Deshalit:53,Talmi:60,Heide:90}, 
	we assume that the proton 
         interacts with the odd neutron through a spin-dependent contact interaction, 
	$V_{np} = F_0 \delta(\mathbf r_1-\mathbf r_2) \left(1-\alpha(1 -\pmb\sigma_1\cdot\pmb\sigma_2)\right)$.
	The two parameters  $F_0$ and $\alpha$ were  determined by fitting the experimental 
	 splittings in $^{12}$B (see Fig. \ref{fig:compare_levels})  and optimising the agreement with the transfer cross sections measured  in $^{10}$Li as discussed below.
The scattering lengths of the resulting  $2^-,1^-$ states  are equal to -19  fm  and -5 fm respectively ($\epsilon_{2^-} \approx$  0.05 MeV, $\epsilon_{1^-} \approx$  0.8 MeV).
For the positive parity doublet one finds instead  ($\epsilon_{1^+} \approx$  0.3 MeV, $\epsilon_{2^+} \approx$  0.6 MeV).

Theory  also predicts  the existence  of a resonant, many-body $\widetilde{5/2^+}$ state with centroid at $\approx3.5$ MeV 
which splits into four states $(\widetilde {5/2}^+ \otimes 1p_{3/2}(\pi))_{1^-,2^-,3^-,4^-}$ spanning the energy interval 2-6 MeV.
In the measured energy interval, the main contribution originates from the  $4^-$ state. 
The $\widetilde{5/2^+}$ resonance state  has a pronounced many-body character. It is 
strongly influenced by configurations including two quadrupole phonons and associated anharmonic effects, similarly to the case of $^{11}$Be previously discussed.


 
Theory also predicts the presence of a $3/2^-$ neutron state, 
which splits into four states ($\widetilde{3/2}^- \otimes 1p_{3/2}(\pi)$)$_{0^+,1^+,2^+,3^+}$ 
with energies within the range 3--6 MeV. 

\begin{figure}
    \centering
     \includegraphics[width=0.9\linewidth,trim=4pt 6pt 1pt 4pt,clip]{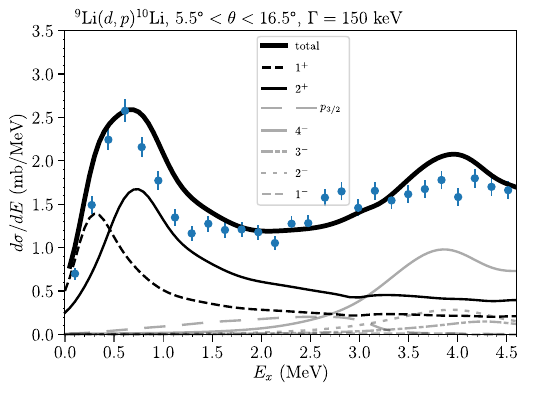}
		\caption{\footnotesize Theoretical prediction (continuous solid curve) of the $^{10}$Li strength function for the $d(^9$Li,$p)^{10}$Li reaction at 100  MeV incident energy and $\theta_{CM}=[5.5^\circ,16.5^\circ]$ in comparison with the experimental data  (solid dots with errors).  The optical potentials used in this calculation were taken from  \cite{Schmitt}. From \cite{10Li}.} 
\label{fig:strength_10li}
	\end{figure}

A good test of the theory is provided by the comparison with the data measured in the d($^9$Li,$^{10}$Li)p reaction at 100  MeV incident energy \cite{Cavallaro:17}, shown in Fig. \ref{fig:strength_10li}, where the computed contributions associated with the different multipolarities are shown.  

\subsection{$^{11}$Li}

The study of the fragile nucleus $^{11}$Li is particularly delicate, in view of the fact that $^{10}$Li is unbound.
From our perspective, the contribution of the induced pairing interaction plays a particularly important role in determining the stability of this system, characterized by  a two-neutron separation energy of only 370 keV. 
Furthermore, the collective modes of the system display a qualitative modification going from $^{9}$Li to $^{11}$Li, because  of the appearance of appreciable low-lying dipole strength in $^{11}$Li, which is not present in the $^9$Li core.

The vibrational quadrupole modes of the $^{9}$Li core are obtained from a p-h RPA calculation making use of a separable quadrupole-quadrupole force, with the radial form factor $r dV/dr$ and with the self-consistent value of the  coupling constant
(see \cite{BohrII}, Eq.(6-74)).
In the calculation we assume an occupation factor equal to 1/4 for the   $1p_{3/2}$ proton level (equal filling approximation). 

The calculation of PVC effects in $^{11}$Li poses a delicate problem, because the presence of two valence neutrons causes a strong modification of the features of the collective vibrations with respect to $^{9}$Li.
In fact $^{11}$Li is well known for the appearance of the so called soft mode, which brings about 8\% of the sum rule to very low energies. This abrupt difference represents a sort of phase transition, with respect to the $^9$Li dipole response, for which such low energy strength does not appear, so that the use of $^9$Li as a dynamical core is not appropriate.
We will then perform a RPA calculation including $s_{1/2}$
and $p_{1/2}$ orbitals lying at the energies of the renormalized  ${\widetilde{1/2^+}}$ and ${\widetilde{1/2^-}}$
states.
The resulting low-lying dipole strength must be taken into account in the calculation of the PVC effects on the valence neutrons, but this in turn requires a careful consideration of the Pauli principle, because a large part of the dipole strength is produced by transitions between the valence neutrons themselves.
However,  the RPA wavefunction of the low-lying dipole contains also components involving  core orbitals,   that can contribute to the PVC between valence neutrons and core. Such components represent strength which is brought to very low-energy by the mixing between the halo dipole excitations and the core dipole excitations lying at 5-10 MeV.
In  Table \ref{tab:RPA_11Li_dipole} the RPA amplitudes of one of the strongest roots contributing, in our discretized continuum calculation,  to the soft mode are presented. One can see that some components involve the halo orbitals $1p_{1/2}$ and the $s_{1/2}$, while  others (including proton transitions, not shown here) involve p-h excitations of the core, and may be excited by PVC. 


A simple estimate of the effect of the elimination of these Pauli-forbidden processes may be obtained considering that the coupling to the dipole mode (see Eq.(\ref{hpvc_rpa})) must be hindered by a factor (1 - $\sum_{valence} [X_{ph} + Y_{ph}]/ \sum_{all} [X_{ph} + Y_{ph}])$, suppressing the contribution of the 
RPA amplitudes of the dipole mode  that involve the halo states. It can be readily checked that this factor is equal to about 0.5 for the dipole mode wave function shown in the table. The contribution to the PVC remains far from negligible, due to the very low energy of the mode.

 In the actual calculation, the Pauli blocking is taken into account by not including the valence components in the evaluation of the transition density entering the evaluation of the $H_{PVC}$ matrix element
 (see Eq. (\ref{trans_rpa})).

\begin{table}[t]
\centering
\small
\setlength{\tabcolsep}{4pt}

\caption{\footnotesize RPA wavefunction of the strongest low-lying dipole vibration of $^{11}$Li ($E_{1^-}=0.75$ MeV),  contributing most importantly to the pairing-induced interaction. All amplitudes refer to neutron transitions. The value $\chi_1 = 0.0043$ MeV$^{-1}$ was used for the isovector coupling constant, yielding good agreement with experiment. This value coincides within 25\% with the self-consistent value of 0.0032 MeV$^{-1}$ (see \cite{BohrII}, Eq. (6-127)) . The resulting strength function integrated up to 4 MeV exhausts 7\% of the Thomas-Reiche-Kuhn energy-weighted sum rule, compared with the experimental value of 8\%.}
\label{tab:RPA_11Li_dipole}

\begin{tabular}{lccccccc}
\hline
 & $\begin{matrix}1p_{1/2}^{-1}\\2s_{1/2}\end{matrix}$
 & $\begin{matrix}1p_{1/2}^{-1}\\3s_{1/2}\end{matrix}$
 & $\begin{matrix}1p_{1/2}^{-1}\\4s_{1/2}\end{matrix}$
 & $\begin{matrix}1p_{1/2}^{-1}\\1d_{3/2}\end{matrix}$
 & $\begin{matrix}1p_{3/2}^{-1}\\5d_{5/2}\end{matrix}$
 & $\begin{matrix}1p_{3/2}^{-1}\\6d_{5/2}\end{matrix}$
 & $\begin{matrix}1p_{3/2}^{-1}\\7d_{5/2}\end{matrix}$ \\
\hline
$X_{ph}$ & 0.847 & -0.335 & 0.244 & 0.165 & 0.197 & 0.201 & 0.157 \\
$Y_{ph}$ & 0.088 & 0.060 & 0.088 & 0.008 & 0.165 & 0.173 & 0.138 \\
\hline
\end{tabular}

\end{table}


\begin{figure}[b!]
    \centering
    \includegraphics[width=0.4\linewidth]{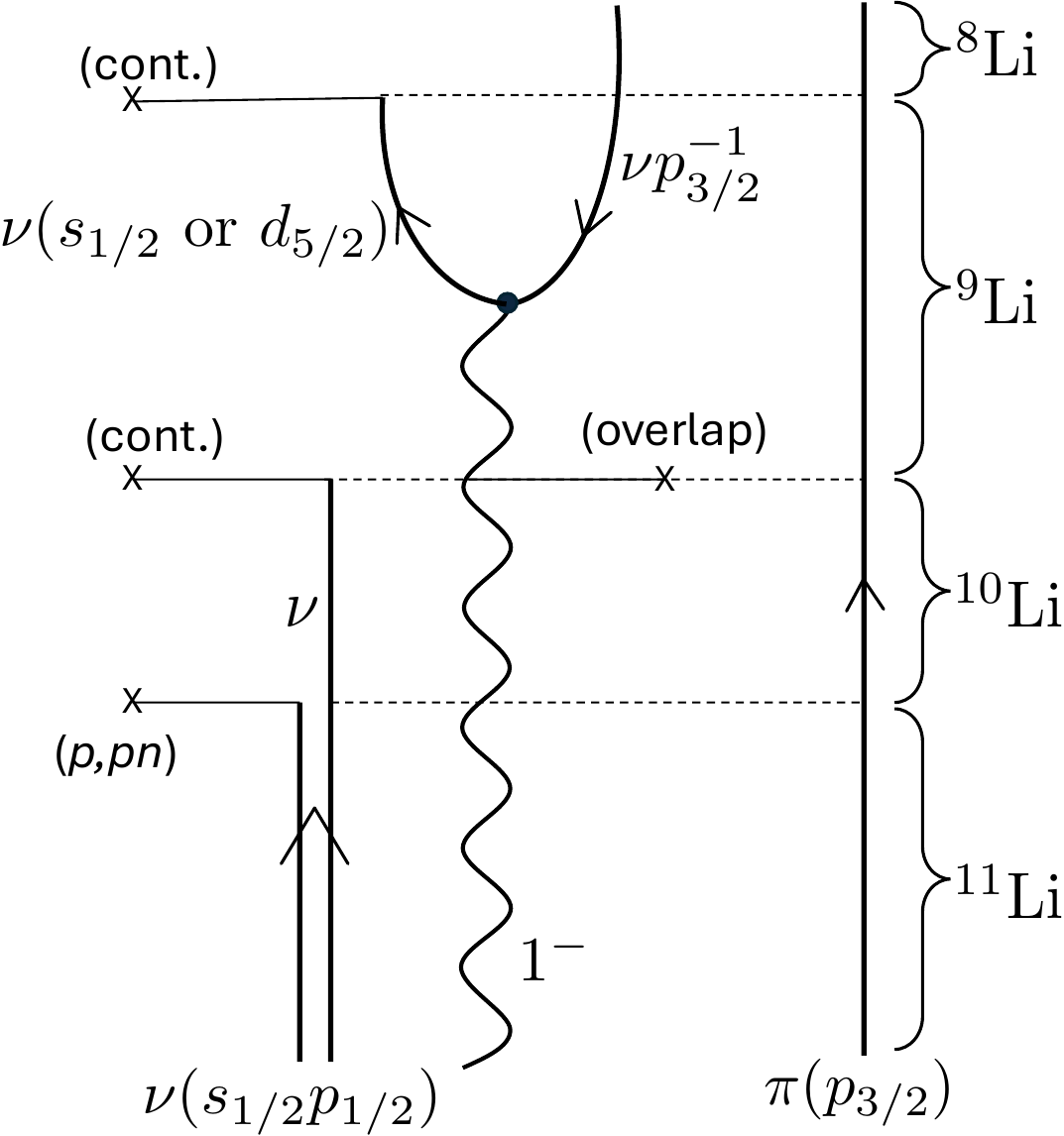}
    \caption{\footnotesize Schematic description of a (p,pn) reaction on $^{11}$Li, probing the dipole component of the halo, and leading to $^{8}$Li via two successive neutron emissions.}
    \label{fig:11Li_ppn}
\end{figure}

The extended pp-RPA matrix is diagonalized using the Argonne $V_{14}$ interaction as the pairing force. 
Neglecting the PVC, no 
bound state is found. Only when core excitation effects are considered a bound state is obtained, with $S_{2n}$ = 0.3 MeV (a similar conclusion was reached in a microscopic six-cluster model of $^{11}$Li \cite{Suzuki}). This agreement with experimental binding energy is a nice feature of the model, but can be obtained in different ways.
For example, adjusting  the mean field potential, a bound system can be obtained by just using the bare interaction, without the need of including any PVC or core excitation. In fact many models in literature adopt this strategy (see below). However, it must be emphasized that the mixing to core excitations represents an  experimental fact, as discussed above in connection with $^{11}$Be. The main virtue of our model of $^{11}$Li does not  consist in  producing an accurate binding energy, but rather in taking into account the couplings that lead to core excitation components in the ground state wave function, and allowing  the interpretation of transfer experiments, as  described below.

\begin{figure}[h!]
        \centering
          \includegraphics[width=0.5\linewidth]{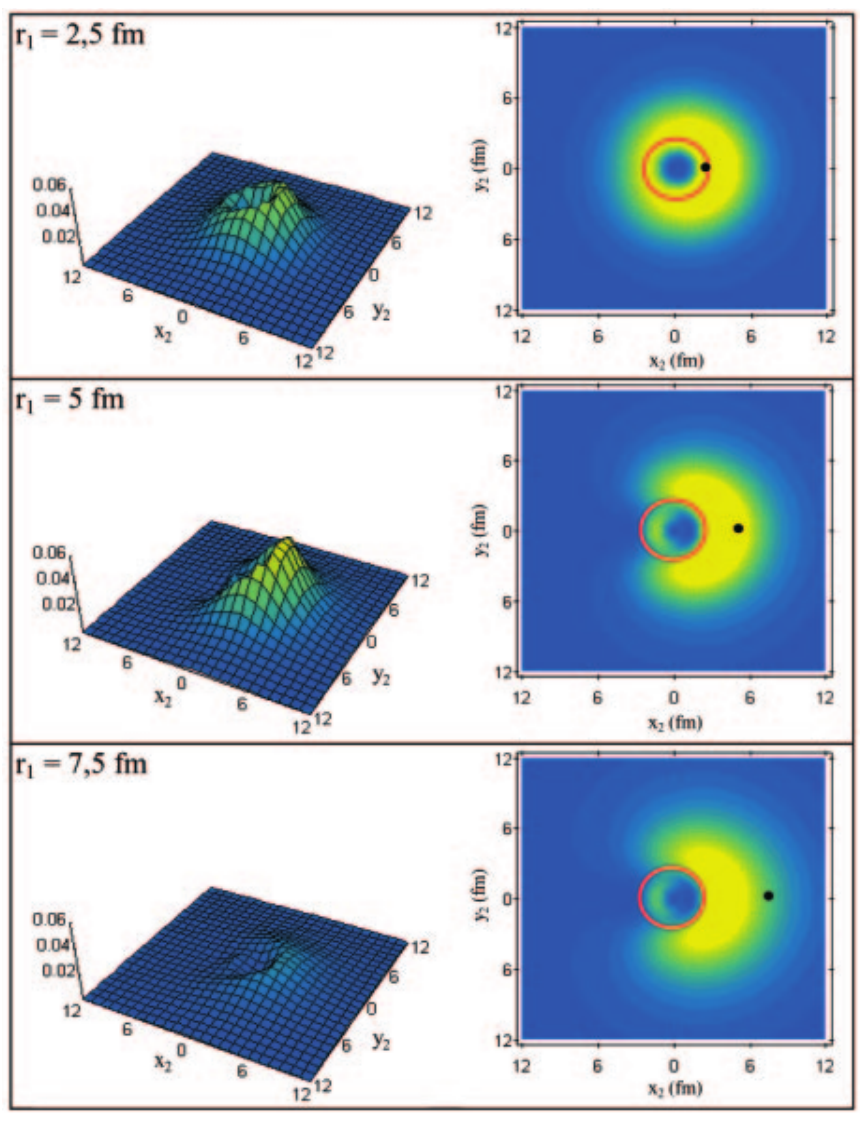}
        \caption{\footnotesize Spatial structure of the two-neutron Cooper pair forming the halo of $^{11}$Li. The modulus squared wave function  describing the motion of the two halo neutrons around the $^9$Li core (normalized to unity and multiplied by $16\pi^2 r_{1}^2 r_{2}^2$
is displayed as a function of the cartesian coordinates $x_2 = r_2 cos(\theta_{12})$ and $y_2 = r_2 sin(\theta_{12})$ of particle 2, for fixed value of the
position of particle 1 ($r_1$ = 2.5, 5, 7.5 fm) represented in the right panels by a solid dot, while the core $^9$Li is shown as a red
circle. The numbers appearing on the $z$-axis of the three-dimensional plots displayed on the left side of the figure are in units of fm$^{-2}$.
From \cite{11Li_struc}.}
    \label{fig:gialla}
\end{figure}

The halo wave function, expanded in a basis of two-particle configurations 
$|(n_1, lj;n_2,lj)_{0^+}\rangle$, and two-particle plus vibration configurations $|(n_1, lj; n_2,l'j')_{L^\pi}\otimes \Phi_{L^\pi} \rangle_{0^+}$  with single particle levels of a  discretized continuum of a box of radius $R_{box} = 40$ fm reads:
\begin{equation}
 \Psi_{11Li} = 0.7 \Psi_{11Li}^{pp} + 0.7 |(p_{1/2} s_{1/2})_{1^-}) \times 1^-\rangle + 0.1  |(s_{1/2} d_{5/2})_{2^+} \times 2^+\rangle
 \label{11Liwf_pp_phon}
\end{equation}
This wavefunction is characterized by  the very large presence of dipole excitation due to the coupling to the core components of the soft mode discussed  above.


The large dipole component might be experimentally revealed by a (p,pN) reaction, as shown very schematically in  Fig. \ref{fig:11Li_ppn}. If a halo neutron is struck by the proton, the remaining $^{10}$Li will decay by nucleon emission leaving the $^{9}$Li in a ${1^-}$ excited configuration, which in turn might decay 
by neutron emission, leading finally to a $^8$Li nucleus. Although this nucleus might be produced by other mechanisms involving the core nucleons, such process is expected to dominate due to large spatial distribution of the halo.

In turn, the pp-component of the wavefunction
shows a high degree of configuration mixing:
    \begin{equation}
     \Psi_{11Li}^{pp} \approx \sqrt {45\%} |(s_{1/2})^2\rangle + \sqrt {45\%} |(p_{1/2})^2\rangle + \sqrt {10\%} |(d_{5/2})^2\rangle
    \label{11Liwf_pp}
    \end{equation}

The induced interaction $V_{ind}$ acts mainly at the surface ($r \approx 3$ fm), while $V_{int}$ acts at short distances. Together, they create a spatial correlation where the two neutrons spend most of their time on the same side of the core (see Fig.  \ref{fig:gialla}). The calculated radii of the neutron halo and of $^{11}$Li are equal to 6.6 fm and to 3.5 fm respectively.

In 2008, Tanihata et al. performed a remarkable experiment at TRIUMF measuring the cross-section of the two-neutron transfer reaction $^{11}${Li}(p,t)$^9${Li} at $E_{lab}$= 33 MeV \cite{Tanihata}. The data revealed a significant population not only of the $^{9}${Li} $3/2^-$  ground state  but also of the first $1/2^-$ excited state lying at 2.69 MeV. This finding challenged simple theoretical descriptions of the $^{11}${Li} halo structure.

\begin{figure}[h!]
    \centering
    \includegraphics[width=0.6\linewidth,trim=4pt 35pt 1pt 50pt,clip]{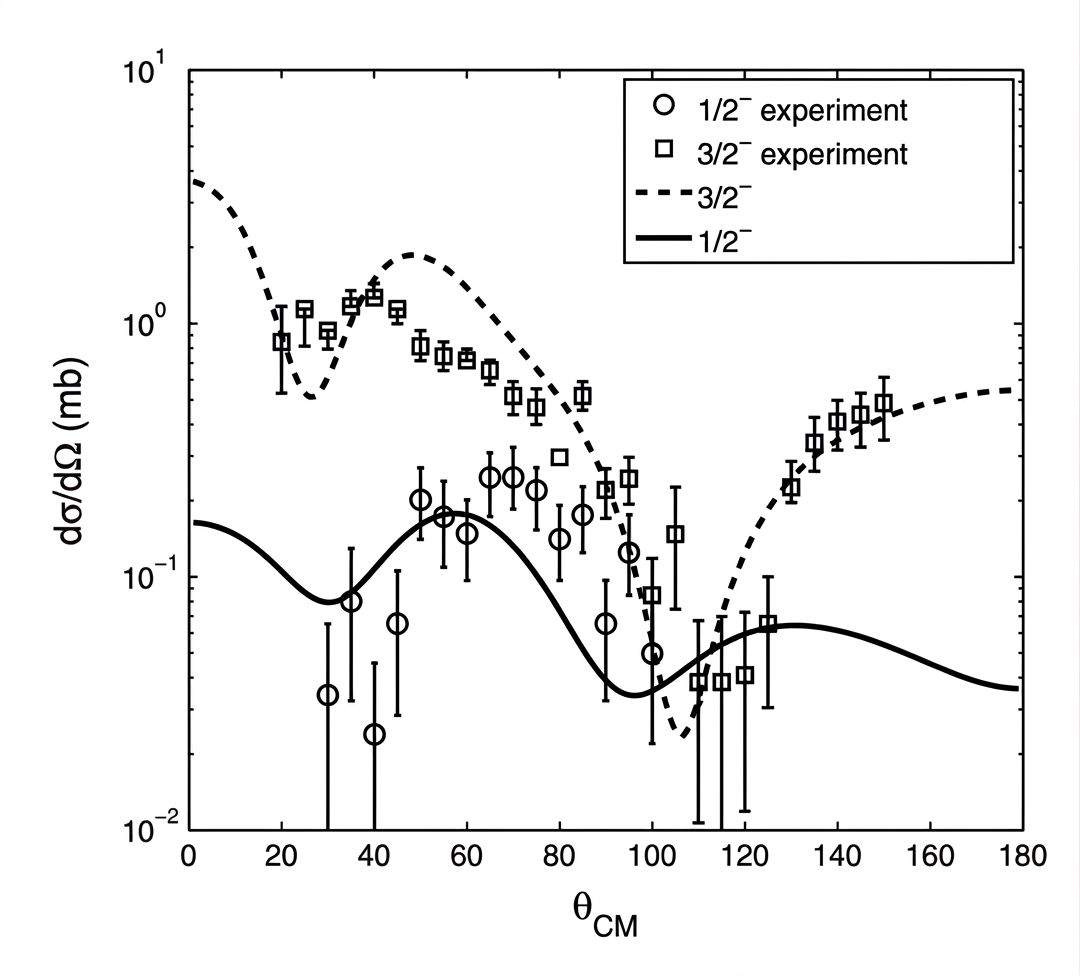}
    \caption{\footnotesize Experimental and total theoretical differential cross
sections (including two-neutron  transfer from the halo as well as other
 channels) of the $^{11}${Li}(p,t)$^9${Li}
reaction at 3 MeV/u in the center of mass frame populating the
ground state ($3/2^-$) (dashed line) and the first excited state ($1/2^-$ ; 2.69 MeV) (solid line) of $^9$Li. The optical potentials used for the calculations were taken from the original experimental reference \cite{Tanihata}, except in the case of the deuteron channel, for which a global parametrization was used \cite{hanDeuteronGlobalOptical2006}. From \cite{Potel_2010}.}
\label{fig:11Li_dwba}
\end{figure}

An interpretation of these data could be provided 
by a second order DWBA calculation using the wavefunctions  (\ref{11Liwf_pp_phon}) and
 (\ref{11Liwf_pp})  (\cite{Potel_2010}, see Fig. \ref{fig:11Li_dwba}). The cross section calculation (see  \cite{Potel_review,Potel_book}) includes the $T$-matrix associated with the simultaneous transfer of two neutrons, 
  \begin{align}
   \nonumber    T^{(1)}(j_i, j_f) &= 2 \sum_{\sigma_1 \sigma_2} \int d\mathbf{r}_{fF}\, d\mathbf{r}_{b1}\, d\mathbf{r}_{A2}\,
  [\Psi_{j_f}(\mathbf{r}_{A1}, \sigma_1) \Psi_{j_f}(\mathbf{r}_{A2}, \sigma_2)]^{0*}_{0}\,
  \chi^{(-)*}_{bB}(\mathbf{r}_{bB})\, v(\mathbf{r}_{b1})\,\\
 &\times [\Psi_{j_i}(\mathbf{r}_{b1}, \sigma_1) \Psi_{j_i}(\mathbf{r}_{b2}, \sigma_2)]^{0}_{0}\,
  \chi^{(+)}_{aA}(\mathbf{r}_{aA})
 \end{align}
 the sequential process (p $\rightarrow$ d $\rightarrow$ t), 
  \begin{align}
  \nonumber T^{(2)}_\mathrm{succ}(j_i, j_f) &= 2
  \sum_{K,M} \sum_{\substack{\sigma_1 \sigma_2 \\ \sigma_1' \sigma_2'}}
  \int d\mathbf{r}_{fF}\, d\mathbf{r}_{b1}\, d\mathbf{r}_{A2}\,
  [\Psi_{j_f}(\mathbf{r}_{A1}, \sigma_1) \Psi_{j_f}(\mathbf{r}_{A2}, \sigma_2)]^{0*}_{0}\,
  \chi^{(-)*}_{bB}(\mathbf{r}_{bB})\, v(\mathbf{r}_{b1})\,\\
  \nonumber &\times[\Psi_{j_f}(\mathbf{r}_{A2}, \sigma_2) \Psi_{j_i}(\mathbf{r}_{b1}, \sigma_1)]_{KM} \int d\mathbf{r}'_{fF}\, d\mathbf{r}'_{b1}\, d\mathbf{r}'_{A2}\,
  G(\mathbf{r}_{fF}, \mathbf{r}'_{fF})\,\\
 &\times [\Psi_{j_f}(\mathbf{r}'_{A2}, \sigma_2') \Psi_{j_i}(\mathbf{r}'_{b1}, \sigma_1')]_{KM}
  \frac{2\mu_{fF}}{\hbar^2}\, v(\mathbf{r}'_{f2})\,
  [\Psi_{j_i}(\mathbf{r}'_{A2}, \sigma_2') \Psi_{j_i}(\mathbf{r}'_{b1}, \sigma_1')]^{0}_{0}\,
  \chi^{(+)}_{aA}(\mathbf{r}'_{aA})
 \end{align}
 as well as the non-orthogonal contribution,
  \begin{align}
  \nonumber T^{(2)}_\mathrm{NO}(j_i, j_f) &= 2
  \sum_{K,M} \sum_{\substack{\sigma_1 \sigma_2 \\ \sigma_1' \sigma_2'}}
  \int d\mathbf{r}_{fF}\, d\mathbf{r}_{b1}\, d\mathbf{r}_{A2}\,
  [\Psi_{j_f}(\mathbf{r}_{A1}, \sigma_1) \Psi_{j_f}(\mathbf{r}_{A2}, \sigma_2)]^{0*}_{0}\,
  \chi^{(-)*}_{bB}(\mathbf{r}_{bB})\, v(\mathbf{r}_{b1})\,\\
 \nonumber & \times [\Psi_{j_f}(\mathbf{r}_{A2}, \sigma_2) \Psi_{j_i}(\mathbf{r}_{b1}, \sigma_1)]_{KM} \int d\mathbf{r}'_{b1}\, d\mathbf{r}'_{A2}\,
  [\Psi_{j_f}(\mathbf{r}'_{A2}, \sigma_2') \Psi_{j_i}(\mathbf{r}'_{b1}, \sigma_1')]_{KM}\\
 & \times 
  [\Psi_{j_i}(\mathbf{r}'_{A2}, \sigma_2') \Psi_{j_i}(\mathbf{r}'_{b1}, \sigma_1')]^{0}_{0}\,
  \chi^{(+)}_{aA}(\mathbf{r}'_{aA}).
 \end{align}

 \begin{figure}[t!]
    \centering
    \includegraphics[width=0.6\linewidth]{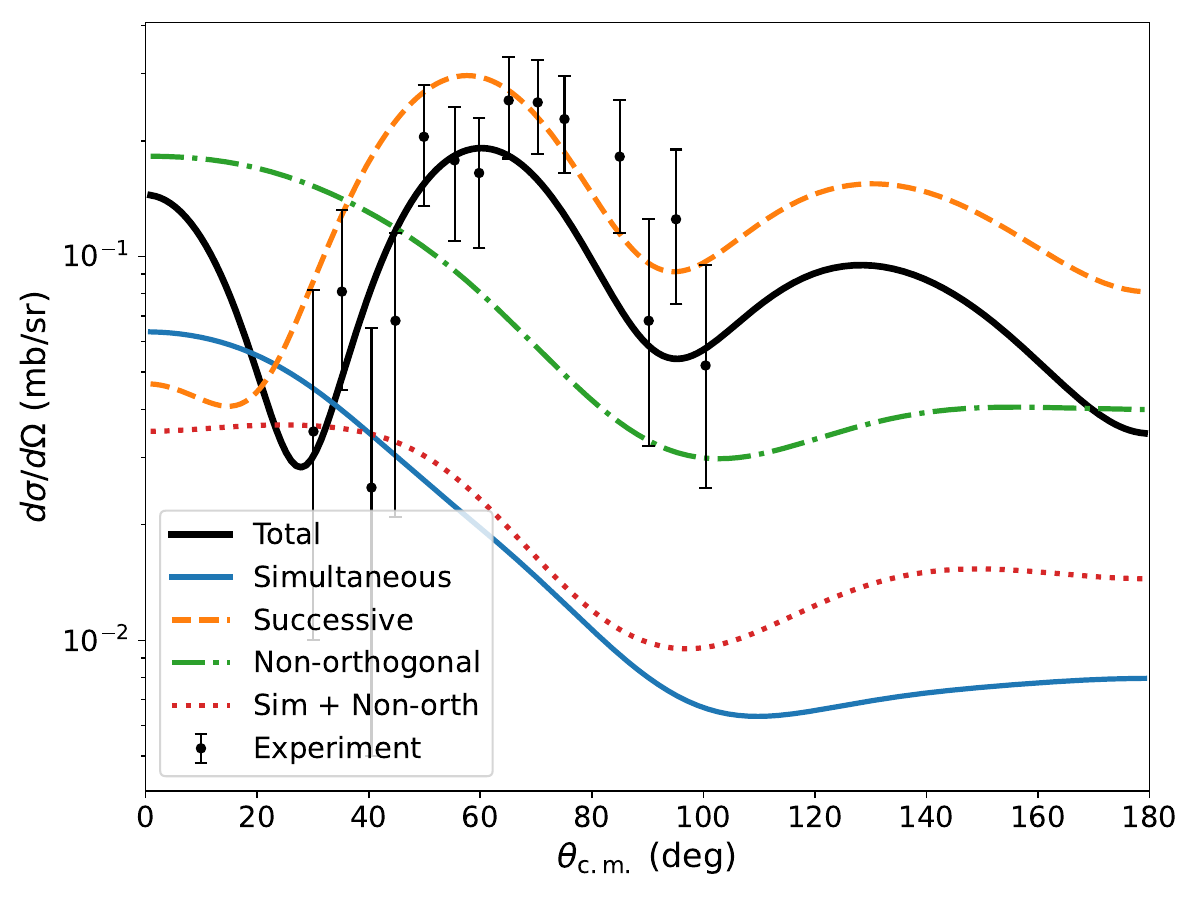}
    \caption{\footnotesize Two-neutron transfer cross section corresponding to the inverse-kinematic reaction $^{11}$Li($p,t$)$^{9}$Li with a 33 MeV $^{11}$Li beam, populating the first ($1/2^{-}; E_x=2.691$ MeV) excited state of $^{9}$Li. The cross sections associated with the terms contributing to the second order prior-prior calculation are shown separately, as well as the result obtained from the partial sum of the simultaneous and non-orthogonal terms (red-dotted curve). The proper coherent sum of the corresponding amplitudes is seen to have a sizable impact both on the magnitude and the shape of the angular differential cross sections. In particular, the conspicuous cancellation of the simultaneous and non-orthogonal terms, which are in relative anti-phase, is observed. The difference with respect to the calculation showed in Fig. \ref{fig:11Li_dwba} is due to the use of a different set of optical potential parameters. This difference was motivated by the need to have a systematic account of the energy dependence of the optical potential for our preliminary calculations at a higher beam energy  (see the caption of Fig. \ref{fig:11Li_66_MeV}).}
\label{fig:11Li_contributions}
\end{figure}

 Although the above expressions correspond to the population of a final state where the two transferred nucleons are coupled to zero angular momentum, they can be generalized for arbitrary final angular momentum transfer. The notation addresses the general 2-step reaction $a(=b+2)+A\to f(=b+1)+F(=A+1)\to b+B(=A+2)$, and the numbers 1, 2, label each one of the transferred neutrons. The successive contribution exhibits the propagation over the intermediate deuteron channels, described by the Green's function $G(\mathbf r_{fF},\mathbf r'_{fF})$.

 It is important to point out that these three terms exhaust all the contributions to the cross section up to second order in the transfer potential $v(\mathbf{r})$. In particular, the non-orthogonal contribution, which shows up as a two-step DWBA term (see \cite{thompson_nunes_2009}) is actually of first order in the transfer potential, and failing to add it coherently to the simultaneous contribution can yield inconsistent results already at first order. An illustration of this fact is that, in the prior-prior representation implemented here, the non-orthogonal and simultaneous contributions have a strong tendency to cancel each other (see Fig. \ref{fig:11Li_contributions}), so that the total cross section is dominated by the successive contribution. In heavy targets, where polarization effects are less important and mean field physics play a more dominant role, this cancellation is essentially complete (see \cite{potel2011calculation}).

\begin{figure}
    \centering
    \includegraphics[width=0.6\linewidth]{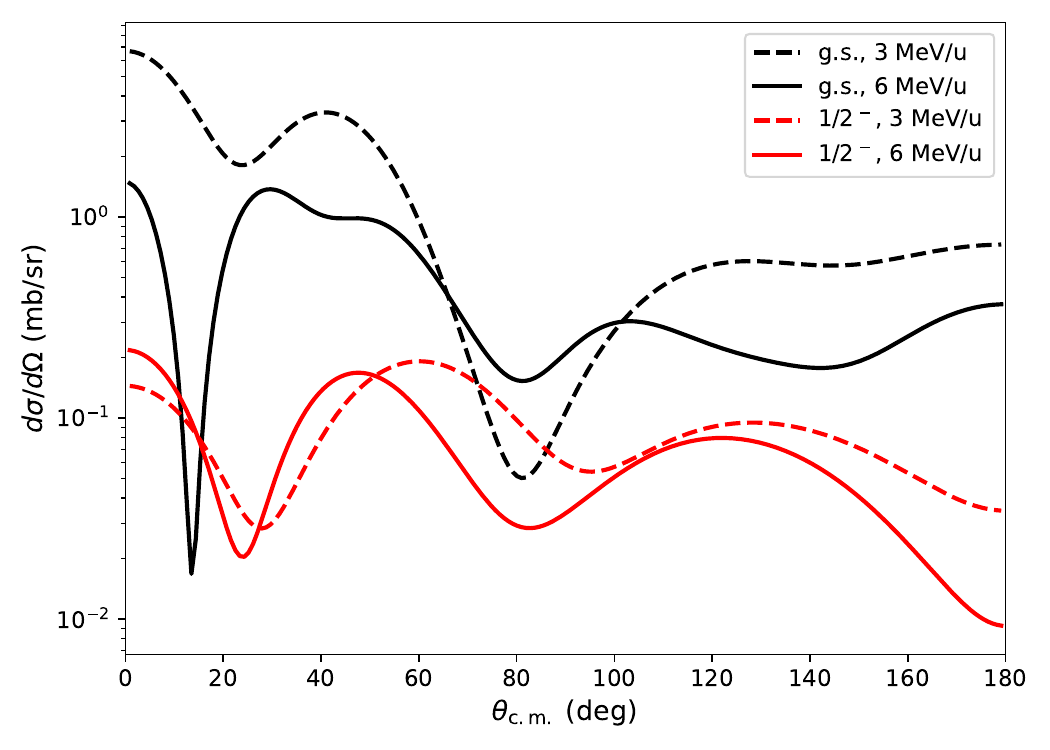}
    \caption{\footnotesize Calculated differential cross sections of the $^{11}${Li}(p,t)$^9${Li}
reaction at 6 MeV/u in the center of mass frame populating the
ground state ($3/2^-$) (black solid line) and the first excited state ($1/2^-$ ; 2.69 MeV) (red solid line) of $^9$Li. We also include the calculations  at 3 MeV/u  for comparison (dashed lines). The energy dependence of the set of optical potentials is consistently accounted for by implementing global parametrizations for the proton, deuteron, and tritium channels (\cite{KoningDelaroche2003,AnCai2006,Pang2009}). Let us point out that these optical potentials are used outside of their range of validity, for lack of global parametrizations dedicated to such a light system. A more reliable calculation beyond this preliminary one would call for a careful choice and calibration of the optical potentials used.}
\label{fig:11Li_66_MeV}
\end{figure}

 According to our calculation, the population of the excited state of the core takes place thanks to the presence of the  $2^+$ admixture in Eq. (\ref{11Liwf_pp_phon}). 
 The magnitude and shape of the angular distribution are in good agreement  with the data 
 (see Fig. \ref{fig:11Li_dwba}). It should be remarked that possible alternative mechanisms for the population of $^9$Li, like direct inelastic excitation and pick-up of a neutron from the $^9$Li core, were estimated and found to be very small. The experiment has recently been repeated at higher energy, as was shown in the talk by X. Wang at this Conference. In Fig. \ref{fig:11Li_66_MeV} we show a preliminary calculation of the DWBA cross sections at the new energy. It is an open question, whether the other mechanisms found to be negligible  at 33 MeV may play a more relevant role at this higher energy.

\begin{figure}[h!]
    \centering
    \includegraphics[width=0.6\linewidth,trim=4pt 10pt 1pt 15pt,clip]{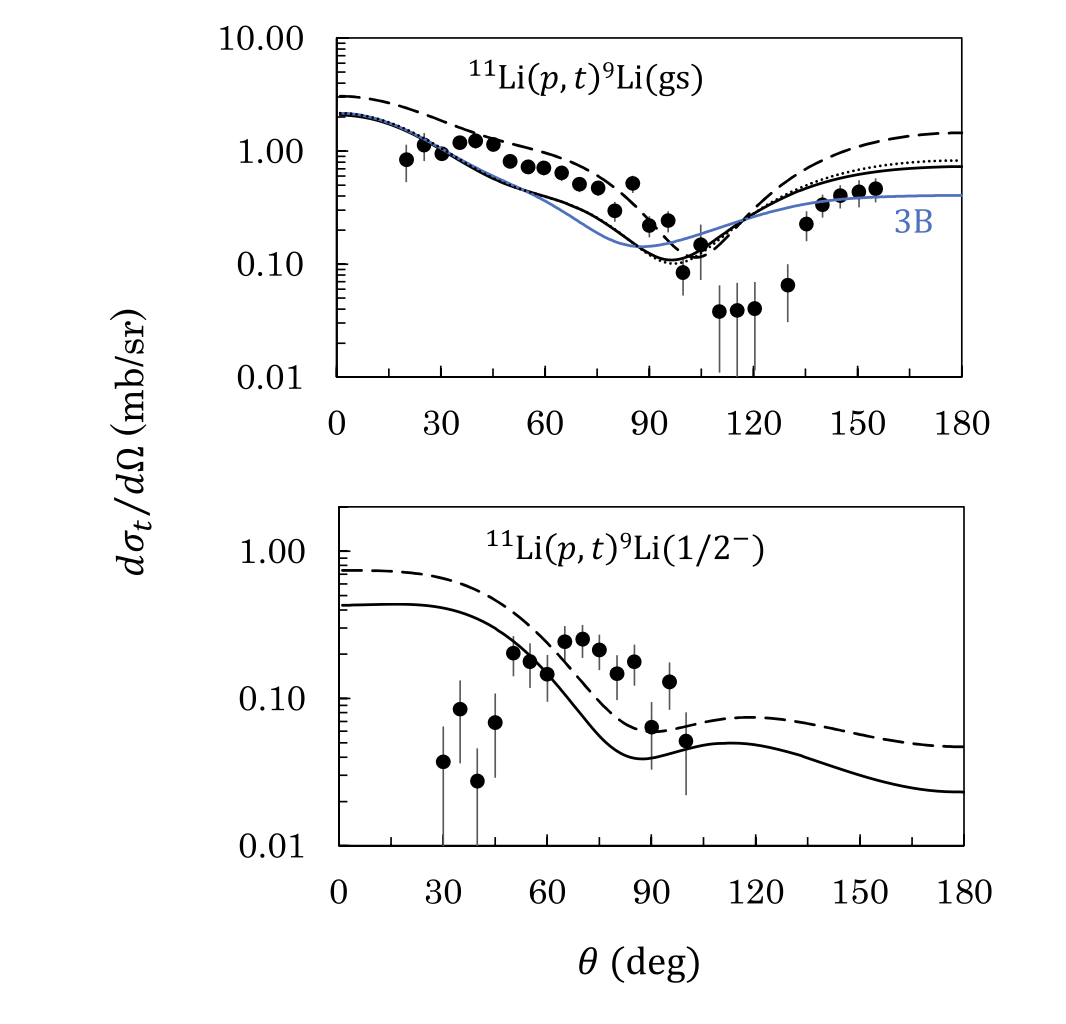}
    \caption{\footnotesize $^{11}$Li(p,t)$^9$Li cross sections with a 33 MeV $^{11}$Li beam, populating the gorund state (top panel) and the first excited state (bottom panel) calculated in \cite{Descouvemont} . 
    Solid and dashed lines are obtained with two different optical potentials. The
dotted line in the top panel corresponds to the $^{11}$Li overlap integral without core excitation. The
curve labeled as '3B' in the bottom panel shows a previous three-body calculation \cite{Descouvemont_previous}.}
    \label{fig:descouvemont}
\end{figure}

Descouvemont \cite{Descouvemont} has also studied the  two-nucleon transfer cross sections populating the ground and excited state of $^9$Li
(see Fig. \ref{fig:descouvemont}). 
The structure calculation is done by a shell model method using the Minnesota potential. In the reaction calculation, only the simultaneous contribution is included in the prior-prior representation. 
The angular distribution for the transfer to the ground state $3/2^{-}$ is similar to our result, although the deep minimum at around 105$^\circ$ is less pronounced.
The effect of core excitation is negligible in this calculation, while in our case the large admixture with the dipole excitation reduces the cross section by approximately a factor 2.
The magnitude of the cross section for the $1/2^-$ excited state is similar 
to ours, although the calculated admixture of the quadrupole core excitation is about 0.1, much larger than our result. 
Possibly, the larger value of the admixture compensate for the lack of the successive contribution.  
Actually, the shape of the angular distribution   is rather poor compared to the data, and 
bears some similarity with  our simultaneous contribution (see Fig. \ref{fig:11Li_contributions}). 

\subsubsection{Alternative approaches}

In the following we  briefly mention some representative model developed to study $^{11}$Li.

Three-body models have been frequently used (see e.g. \cite{Sagawa2005,Fortunato2014,Gomez,Singh2022} for a few representative works)  to study the spectrum and the reactions of two-neutron halos, following the seminal work by Bertsch and Esbensen \cite{Bertsch}. In this approach the core dynamics is frozen, and phenomenological, parity-dependent static  potentials between the core and the valence neutrons  are introduced to reproduce the experimental parity inversion. Furthermore,  a density-dependent pairing interaction is introduced in such a way to reproduce the experimental two-neutron separation energy of the system.  In this way, a good reproduction of the experimental radii,  low-lying excitations and n-n correlations \cite{Kubota,casal_corr} can be obtained, also in other two-neutron halos.  A comparison with the  spatial correlations  calculated including PVC and previously  shown in Fig. \ref{fig:gialla} would be interesting but it  has not been attempted. 

The work of \cite{Meng1996,Meng1998} applied the Relativistic Hartree-Bogoliubov (RHB) theory with the NL3 interaction to provide a microscopic description of this phenomenon, focusing on the role of the pairing correlation and the contribution of the continuum. The $1p_{1/2}$ and $1s_{1/2}$ orbitals turn out to be much closer to each other than in non relativistic calculations, so that a superfluid solution can  be obtained with the Gogny interaction. The pairing force scatters neutron pairs from the $1p_{1/2}$ shell into the $2s_{1/2}$ shell in the continuum,  and the resulting two-neutron separation energy and radii are in very good agreement with experimental data. 
This is a pure mean field approach, and consequently
there is no explicit connection with the quadrupole vibrational modes of the final $^9$Li that plays such an important role in our model. 
As far as we know, this approach has not been used to interpret reaction data. 

\begin{figure}
    \centering
    \includegraphics[width=0.7\linewidth]{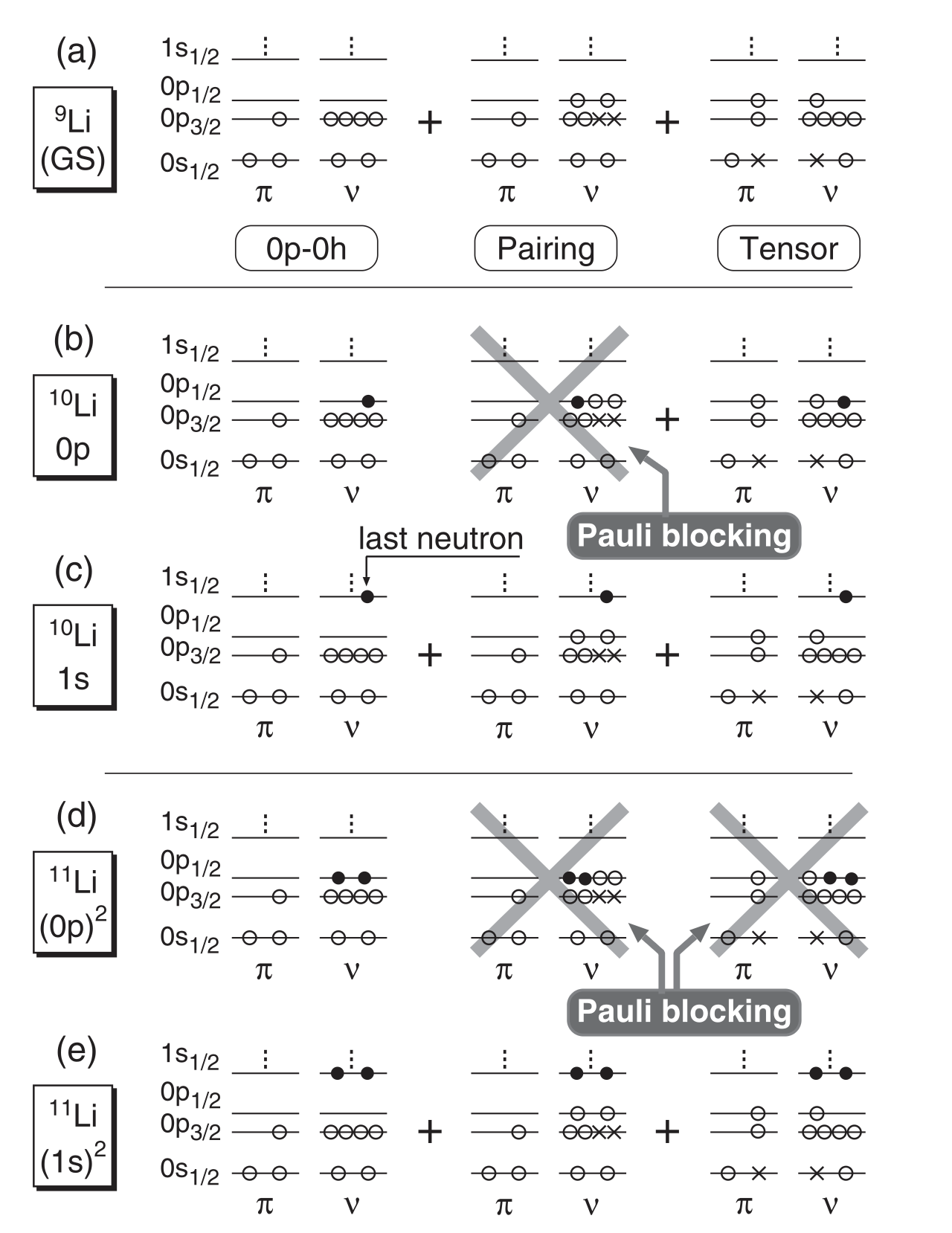}
    \caption{\footnotesize Schematic illustration of  Pauli-blocking  in $^{10}$Li and $^{11}$Li, according to \cite{Myo}.}
    \label{fig:myo}
\end{figure}

The {Tensor-Optimized Shell Model} has been  systematically applied to analyze light systems and in particular $^{9,10,11}$Li. In this approach  the parity inversion in $^{10}$Li and $^{11}$Be  is caused by  the "Pauli blocking" of tensor and pairing correlations in the $^{9}$Li and $^{10}$Be cores by the presence of additional valence neutrons \cite{Myo} (see Fig. \ref{fig:myo}). This blocking effect raises the energy of $p$-shell configurations, effectively reducing the $s-p$ shell gap and allowing for parity inversion and for the halo formation in $^{11}$Li and the $s-p$ inversion in $^{10}$Li. The $s^2$ component of the valence neutrons in the wavefunction of $^{11}$Li increases to approximately 50\%  when both tensor and pairing correlations are included. This high $s$-wave probability is essential for reproducing the halo structure and the large matter radius of $^{11}$Li.
The blocking of ground state correlations as a key factor to explain the parity inversion is conceptually analogous to our approach, in which however quadrupole correlations (which are very sensitive to the parity of the state)   instead of tensor correlations,   play the main role.

As far as we know, the interpretation of transfer experiments populating excited states of the core has not been considered by the models mentioned in this subsection. It is an open question, whether such experimental results can be reproduced without  introducing the surface fluctuations of the core.

\subsection{$^{14}${C}}

Contrary to the previous examples, $^{14}$C is a well-bound nucleus. 
Our main interest  lies in the calculation of the particle-particle strength (see Eq. \ref{eq:strength}).
The existence of low-lying 
pairing vibrations (PV)  involving  particle-particle (pp) or hole-hole (hh) excitations has  been established  for a long time in several mass regions, mostly by  two-nucleon transfer reactions.
 The existence of high-lying PV excitations, popularly known as Giant Pairing Vibrations (GPV), with energies of the order of 2$\hbar \Omega$ ($\hbar \Omega \approx 41/A^{1/3}$ MeV denotes the distance between major  shells in spherical nuclei)  was proposed theoretically long ago, on the basis 
of schematic calculations within pp-RPA
 \cite{BrogliaandBes1977,Bort2016}. These modes have then been studied within bound representations \cite{Herzog, Fortunato}
and taking the continuum into account within the shell model in the Bergreen
representation  \cite{Liotta}
and  within continuum RPA calculations \cite{Khan,Avez,Matsuo}. 
The GPV has not been clearly identified, in spite of many experimental attempts  carried out in different mass regions and with different probes (see e.g. \cite{Assie,Laskin} and refs. therein). 
 Recent transfer experiments between heavy ions populating  $^{14}$C and $^{15}$C  \cite{Cappuzzello_2015,Cappuzzello_epj,Bonaccorso}, however,  have identified bumps 
which have been proposed to be  a signature of the GPV.

As an addition mode, particularly in light nuclei, the GPV should typically lie  close to the single-particle continuum, so it may be seen as an excited two neutron quasi-halo structure, despite the fact that the  low lying ground state PV is well bound. Moreover, at such energies, another 'continuum' appears, namely that associated with the $pp \otimes b$ configurations, to which the GPV would couple on-shell through $H_{PVC}$. For these reasons, the methods presented above for halo nuclei appear particularly  suitable for the study of the GPV. The only difference with respect to the description of previous sections is that we will be dealing not only with the ground state solution of the A+2 Hamiltonian (see Table \ref{Table_Ext_RPA}), but  with the whole excitation function in the continuum, where the existence  of  the  pp-resonance should be revealed. This is done by analyzing the so called strength function introduced in Eq. (\ref{eq:strength}).
A  refined description of the coupling to the unbound configurations appearing in  the $H_{2\nu}$ Hamiltonian is then required. This has been achieved, as was already mentioned above, by repeating the calculations in different boxes and averaging over the resulting strength functions.

 As for the bare pairing interaction, $V_{int}$, we use the Gogny interaction reducing its strength by 20\%, in order leave room for the contribution of the induced pairing interaction to the pairing gap \cite{Schuck,Pastore,Idini}. In fact the Gogny interaction was devised to reproduce the pairing gaps deduced from the empirical odd-even mass difference. 
 The bare Woods-Saxon potential has been obtained as in the previous Sections (see Table \ref{bare_para}), except that a constant effective mass
equal to the reduced mass, $m^* = 0.92 m$ has been used, leading to the parameters $V_{WS}= 72.0$ MeV, $V_{ls}=$ 28.5 MeV, $a_{WS}= 0.64$ fm, $R_{WS}=$ 2.27 fm.
This potential 
 produces two  deeply bound neutron hole states ($e_{1s1/2}$ = -34.0 MeV,
$e_{1p3/2}$ = -15.5 MeV). As for particle states, one finds a  $1p_{1/2}$ orbital lying at  - 5.8 MeV  and a weakly bound $2s_{1/2}$ orbital
lying at $e_{2s1/2}= $ -0.4 MeV.
Furthermore,  one finds a pronounced $d_{5/2}$  resonance at 0.8 MeV.  On the other hand,  there is no low-lying single-particle 
$d_{3/2}$ resonance, in keeping with the experimental data \cite{Ohnuma,Tanifuji}. 
Including the coupling with the $2^+$ low-lying vibrational state of the core, 
using  the values $\hbar \omega_{2^+} = 4.44$ MeV and $\beta_2= 0.46$ 
leads to three renormalised  many-body  particle states with energies  ${\tilde e}_{1/2^-} = -5.0,  {\tilde e}_{1/2^+} $= -2.0  and ${\tilde e}_{5/2^+}$ =  -1.4 MeV,
to be compared with the experimental values  -4.9, -1.9 and -1.1 MeV   respectively. 
The effect of PVC on these levels is analogous to that previously discussed  in $^{11}$Be.
Note in particular the strong repulsive Pauli blocking effect on the $1p_{1/2}$ orbital.
We found that the effect of the coupling to the lowest octupole mode is negligible. 

\begin{figure}[t!]
\center
\includegraphics[width=0.8\textwidth]{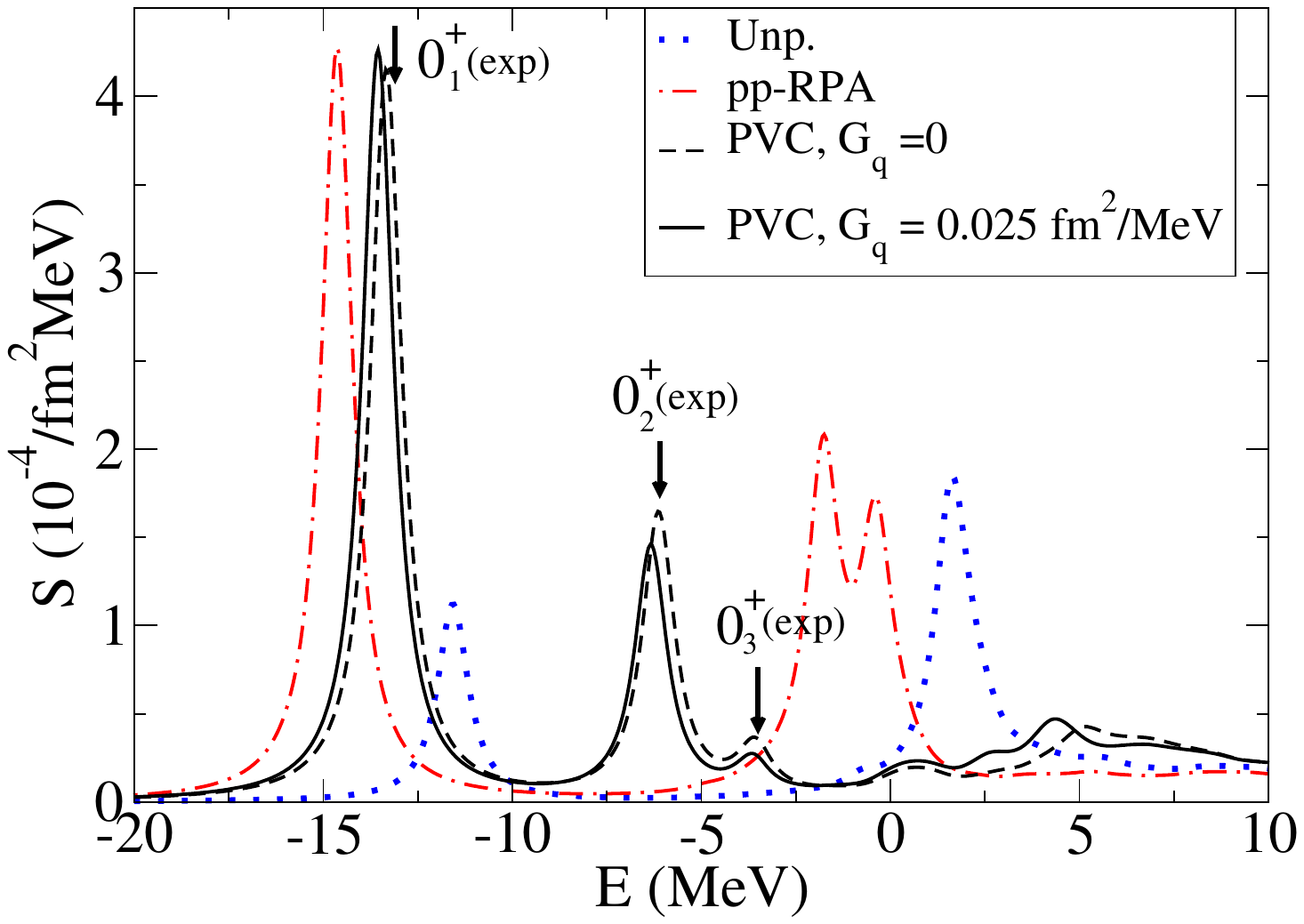}
\caption{\footnotesize  Monopolar  strength functions in $^{14}$C, calculated as discussed in the text.
The strength functions have been averaged by a Lorentzian of FWHM =1 MeV for graphical convenience.
The experimental  energy $-S_{2n} $ of $^{14}$C  is indicated by the  arrow $0^+_1$.  
The positions of the two lowest excited $0^+$ states   obtained from their respective  experimental
 excitation energies  are indicated by the arrows $0^+_2$ and $0^+_3$. From \cite{prl_gpv}.
}
\label{fig:strength_14c}
\end{figure} 

In Fig.  \ref{fig:strength_14c} we show the monopolar strength functions  obtained  from the unperturbed mean field 
calculation (curve labeled Unp.) and from  the pp-RPA result using the Gogny interaction and ignoring $H_{PVC}$ (curve labeled pp-RPA).  
We then include the PVC coupling, studying the effects of  
 a quadrupole pairing interaction ($V_{int}^{quad}$) acting among valence neutrons  coupled to $2^+$:
\begin{equation}
V_{int}^{quad} = -\frac{\pi G_q}{5} \sum_{\mu} P_{2\mu}^{\dagger} P_{2\mu} 
\end{equation}
where
\begin{equation}
P_{2\mu}^{\dagger} = \sum_{j_1 j_2} \langle j_1 || dV/dr Y_2 || j_2 \rangle [a_{j_1}^{\dagger} a_{j_2}^{\dagger}]_{2\mu}.
\end{equation}
and we use the  coupling strength $G_q$=0.047 fm$^2$ MeV$^{-1}$.

  The unperturbed strength is dominated by the lowest $(1p_{1/2})^2$ peak lying at $E \approx  -12$ MeV  
  and by the $(1d_{5/2})^2$  peak at $E \approx 2$  MeV.
  The strength associated with the $(2s_{1/2})^2$ lying at $E= - 0.8$ MeV  is instead barely visible in  Fig.  \ref{fig:strength_14c}, because 
  the low-lying $2s_{1/2}$ orbital extends well beyond the nuclear surface and has a small overlap with the form factor.

The pp-RPA calculation lowers the energy of the ground state by about 4 MeV.
The associated pairing strength increases by a factor of about 4.
These features are those expected for a collective  PV, although the number of bound single-particle
states is very limited,
and are associated with  the presence of  a $(1d_{5/2})^2$ component (about 6\%)
and  of backward ($1p_{3/2})^2$ components  (about 4\%) on top of the 
unperturbed $(1p_{1/2})^2$ strength. In addition,
the pairing interaction shifts the  unperturbed sd bump concentrated between -1  and   +2 MeV 
down by about 4 MeV, increasing considerably its total strength.  The resulting correlated bump 
represents a collective excitation in the shell next to the ground state PV
and must be identified with the  GPV,  often discussed in more schematic models.
Comparing the pp-RPA and the PVC calculations, one can observe that  the  PVC increases the  energy of the ground state 
by almost 2 MeV, in keeping with the upward shift associated with the 1$p_{1/2}$ orbital  
in a similar way to that discussed above in connection with  $^{11}$Be. 
We remark that this is the 
main renormalization effect associated with the $ph \otimes b$  sector of our Hamiltonian.
The absolute value of $E(0^+_1)$   (13.5 MeV) is in good agreement  with the experimental two-neutron separation energy  (13.1 MeV).
 The  ground state wavefunction 
 now contains a 13\%  $pp' \otimes 2^+$ admixture, that 
 could be  tested
by the inverse transfer reaction  populating the $2^+$ state in $^{12}$C, similarly to the case of the reaction $^{11}$Li(p,t)$^9$Li$^*$ \cite{Tanihata}.

 Furthermore, the PVC splits  the GPV bump described above.
 Most of the  strength is shifted to lower energy, due to the action of  the polarization diagram on 
 the  2$s_{1/2}$ and 1$d_{5/2}$ single-particle states, 
as well as to the  effect of the induced interaction.
As a result  one finds  two excited  states $0^+_2$ and $0^+_3$
 lying at $E \approx$  -6.3 MeV ($E^* \approx $ 7.2 MeV), with dominant $(1d_{5/2})^2$ and $(2s_{1/2})^2$ components, 
and  at  $E \approx -3.8 $ MeV ($E^* \approx$ 9.7 MeV), also mostly of sd character. The  $pp' \otimes 2^+$  admixtures in $0^+_2$ and $0^+_3$  
are 37\% and 33\% . 
In the recent $^{12}$C($^{18}$O, $^{16}$O)$^{14}$C transfer experiment, only the 
$0^+_3$ state was weakly populated at 
$E_{lab}$ = 84 MeV, while neither the $0^+_2$ nor the $0^+_3$ state were observed at $E_{lab}$ = 275 MeV [6, 7].
These two  states have been populated  in (t,p) reactions \cite{Mordechai} and the ratios of the measured cross sections relative to the ground state are in fair agreement with our calculated strength function.  We have also analyzed the transition densities associated with the $0^+_1,  0^+_2$ and $0^+_3$  states. It is found that $0^+_1$ and $0^+_2$ states  show a rather strong peak and no nodes in the surface region,  in keeping with their collective character. This confirms that the first excited  state deserves  to be considered as a high-lying pairing vibrational state that resembles the prediction of Bes and Broglia formulated long time ago \cite{BrogliaandBes1977}, although  the scarcity of levels in this light nucleus makes that correspondence very schematic.
Instead, the second excited state show a much smaller amplitude and also nodal points, both related to the dephasing of the (sd) components.

Considering now the GPV strength shifted to higher energy, 
the PVC produces a bump in the continuum, which is located in the  region $E^* \approx$ 16-20 MeV, 
not far from the bump detected in the $^{12}$C($^{18}$O, $^{16}$O)$^{14}$C experiment. The width of this bump is
mainly determined by the energy distribution of the most relevant $((pp')_{2^+} \times 2^+)_{J=0^+}$ configurations (with $p,p'= s_{1/2}$ or $d_{5/2}$)
which interact with the $(s_{1/2})^2, (d_{5/2})^2$  configurations  via the PVC and attract some of their pair strength.

To which extent the experimental bump found in \cite{Cappuzzello_2015} may be identified with the bump found in our calculation is an open question which will require a quantitative calculation of the absolute cross section using our wavefunctions. This work is in progress.
In any case our results show that the GPV may be greatly affected both in position and  width/breaking of its strength by the coupling to quadrupole excitations of the core.

\section{Conclusions}
We have reviewed several aspects of the physics of light weakly bound nuclei, focusing on the effects of the coupling between valence nucleons and collective vibrations of the core, which renormalize both spectra and cross sections. Although such effects can sometimes be reproduced within frozen-core calculations through the introduction of suitable effective parameters, we have emphasized a number of features that point to the advantages of a dynamical description. Microscopic energy surfaces in the $\beta,\gamma$ plane exhibit a very shallow character, in contrast with the assumption of a rigidly deformed core.
The NFT approach we have adopted, based on a spherical dynamical core, provides  through the concepts of self-energy and induced pairing interaction a  unified and consistent framework for the analysis of phenomena such as the parity inversion between $1/2^+$ and $1/2^-$ states in $^{11}$Be, the stability of $^{11}$Li and   the cross sections for the population of the $2^+$ excited state of the core  in one- and two-nucleon transfer  reactions. NFT is also suitable for the analysis of resonant states, including PVC effects. The latter may substantially modify the 
profile of strength functions, as compared to mean-field calculations, as was shown in the case of $^{10}$Li and $^{14}$C. 

Finally, we have suggested some new possible developments of the model.
Concerning the study of the spectrum of $^{10}$Be (see Section 4.1.1),  it is expected that the PVC leads to  the splitting of the two-phonon configuration  $2^+ \otimes 2^+$ into the three observed states with $J$ =0,2 and 4,  arising from  due to PVC,  without invoking the rotational picture. We have suggested (see Section 4.3) that the decay of $^9$Li following a $^{11}$Li(p,pn) reaction can give relevant information about the admixture of the dipole component in the wave function of $^{11}$Li, as predicted by our model. Finally (see Section 4.1.1) the contribution of phonon exchange processes can play a major role in the population of excited states of $^{12}$Be in $^{11}$Be(d,p) reactions, which is substantially underestimated in models that neglect the  dynamics of the core.

\section{Acknowledgments}
F. B. acknowledges the I+D+i project with Ref. PID2020-114687 GB-I00, funded by MCIN/AEI/10.13039/
501100011033. This work was supported by the Spanish Ministry of Science, Innovation and Universities (MCIN/AEI/10.13039/501100011033) and by the European Union under project 0787159367-159367-4-824 (NUSTRAP).
\bibliographystyle{unsrt}
\bibliography{references_complete}
\end{document}